\documentclass[prl,aps,amsfonts,amsmath,amssymb,twocolumn,superscriptaddress]{revtex4-2}

\usepackage[dvipdfmx]{graphicx}
\usepackage{bm,ulem}
\usepackage{hyperref}
\usepackage{IEEEtrantools}
\usepackage{pgffor}
\usepackage{pdfpages}
\usepackage{diagbox}
\usepackage{lineno}

\usepackage{mathrsfs}
\usepackage{natbib}
\usepackage{physics}
\usepackage{float}
\usepackage{siunitx}
\usepackage{xcolor}
\usepackage{comment}
\hypersetup{
setpagesize=false,
colorlinks=true,
linkcolor=blue,
citecolor=red,
}

\makeatletter
\AtBeginDocument{\let\LS@rot\@undefined}
\makeatother

\def\supplementfilename{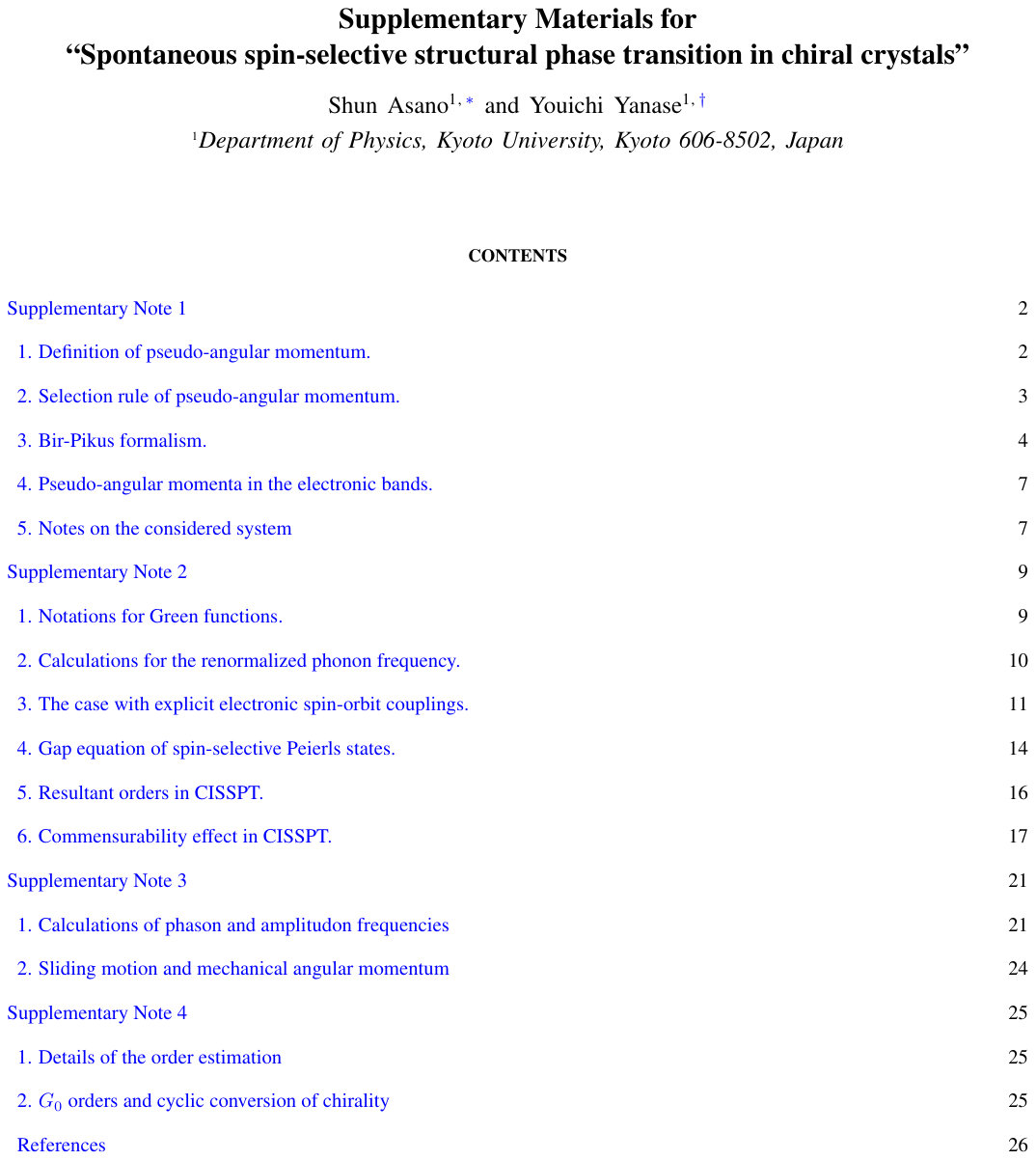}

\pdfximage{\supplementfilename}
\def\numbersupplementpages{\the\pdflastximagepages}

\newif\ifarXiv
\arXivtrue

\restylefloat{table}

\begin{document}

\title{Spontaneous spin-selective structural phase transition in chiral crystals}

\author{Shun Asano}
\email{asano.shun.57x@st.kyoto-u.ac.jp}
\affiliation{Department of Physics, Kyoto University, Kyoto 606-8502, Japan}

\author{Youichi Yanase}
\email{yanase@scphys.kyoto-u.ac.jp}
\affiliation{Department of Physics, Kyoto University, Kyoto 606-8502, Japan}

\begin{abstract}
In this Letter, we predict a structural phase transition unique to chiral crystals with screw symmetry. In chiral crystals, the phonon frequency renormalized by the electron-phonon coupling depends on the handedness of circular polarization. Consequently, the soft mode encoding phonon angular momentum induces spin-selective Peierls gaps in the electronic band, entailing a helical spin density wave and chiral lattice distortion. We also elucidate the chiral signatures and functional implications of collective modes. Our findings offer crucial insights into the emergence of chirality and highlight novel functional aspects of chiral materials and their design strategy.
\end{abstract}

\maketitle


\textit{Introduction}.---
Structure and functionality are inextricably linked in general. In condensed matter physics, the crystal structure rules static and dynamic properties, shaping phenomena ranging from macroscopic phases to microscopic states. This indicates that exotic behaviors can be observed in distinctive structures. A chiral structure, lacking both inversion and mirror symmetry, is the ultimate realization of such a unique system. Recent vibrant research has uncovered remarkable phenomena linked to structural chirality, e.g., chirality-induced spin selectivity (CISS) and magnetochiral anisotropy~\cite{Naaman2022AnnualRevBiophys,Bloom2024ChemRev,Rikken2000Nature,Nagaosa2024AnnualRevCondMat,Yan2024AnnualRevMatRes,Yang2021NatRevChem,Akito2020prl}. Their exotic electronic functionalities offer new avenues for advanced electronics and quantum devices.

Conversely, a complementary line of research based on the opposite perspective has recently emerged: leveraging the functionalities to engineer chiral structures from achiral precursors~\cite{Bousquet2025JPhys,Hayashida2022ACS,Romao2024ACSnano,Zheng2025Science,Zhang2024npj,Fava2025PRL}. Namely, a desirable chiral structure could be induced from an achiral crystal by some external means. Phonons in chiral crystals, so-called chiral phonons, acquire left-handed (LH) or right-handed (RH) circular polarization depending on the chirality of the crystal, since phonons are tightly interlinked with the entire structure~\cite{Zhang2014PRL,Hanyu2018Science,Wang2024NanoLett,Ishito2023NatPhys,Ueda2023Nature,Chen2022NanoLett,Kishine2020PRL,Juraschek2025NatPhys}. Therefore, for example, structural chirality has been revealed to arise by breaking the compensation of phonon handedness, which is inherently degenerate in the parent achiral phase~\cite{Romao2024ACSnano,Zheng2025Science,Zhang2024npj,Luo2023PRX}. These approaches to achiral-to-chiral phase transitions would provide key insights into the ubiquitous formation process of chiral structures in nature, as well as of chiral inorganic crystals~\cite{Budin2010AnnualRevBiophys,Michaeli2016ChemSocRev,Vieda2005PRL,Bailey1998Science,Takano2007EPSL}.

The structural chirality affects the constituents and vise versa, which has been demonstrated independently so far. This raises a fundamental question: Can the interplay between structural chirality and its constituents trigger novel cooperative phenomena? If so, such phenomena could foster both practical advancements and fundamental insights. In conventional crystals, the Peierls transition serves as a classical example arising from the cooperation of the electrons and the lattice structure~\cite{Gruner1988RevModPhys,Gruner1994,Monceau2012AdvPhys,Gruner1994RevModPhys}. The essence of Peierls transitions embodies both elements, giving rise to the charge density wave (CDW) and the lattice deformation simultaneously. Extending this concept to chiral materials, where a unique structure and exotic electronic behaviors coexist, suggests the potential for entirely novel phenomena. Nevertheless, studies on chirality in the Peierls transition have been confined to achiral-to-chiral phase transitions; accordingly, the above question remains unresolved.

\begin{figure*}[t!]
		\includegraphics[width=0.9\linewidth]{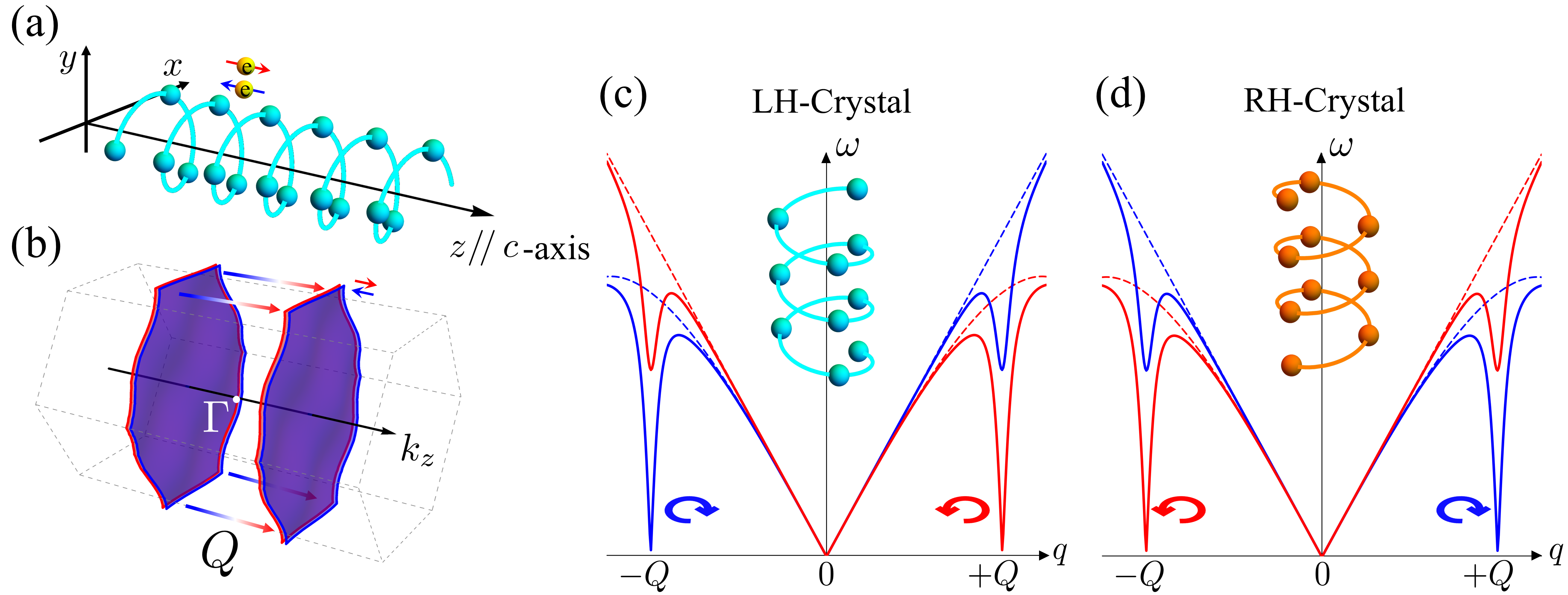}
        \caption{Target chiral materials and Kohn anomaly of chiral phonons. (a) Schematic of the target chiral crystal with screw or helical structure. The $z$ axis is taken along the chiral axis. (b) Schematic electronic structure showing nested spin-up and spin-down Fermi surfaces, which is degenerate in the absence of SOC. (c),(d) TA phonon dispersions for LH and RH crystals, respectively. Dotted and solid lines denote phonon frequencies at high temperatures and just above the transition temperature. Red (blue) curves represent phonons with angular momentum $l^{\mathrm{ph}}_z=+1$ ($-1$).
}
		\label{fig1}
\end{figure*}

Here, we address this question and predict a spontaneous phase transition dubbed the chirality-induced spin-selective Peierls transition (CISSPT). We focus on the case where the parent phase is a chiral metal and consider electron-phonon coupling (EPC) under a screw symmetry as shown in Fig.~\ref{fig1}. Then, we demonstrate the Kohn anomaly of chiral phonons renormalized by electrons, which reveals that the nesting vector involves not merely the linear momentum but also the phonon angular momentum (AM). As a result, a spin-selective Peierls gap is introduced in the electronic band, regardless of whether the original band is spin degenerate or not. Furthermore, we explore the electronic order, lattice distortion, and collective excitation associated with CISSPT.

\textit{Formulations}.---
Although realistic materials are three-dimensional, some materials are highly conductive only in a certain direction, the so-called quasi-one-dimensional (quasi-1D) conductors. We here focus on chiral inorganic or organic conductors with a helical structure, i.e., belonging to chiral space groups. In the main text, as an example, we consider quasi-1D systems with three-fold screw symmetry $\hat{S}_3$. As depicted in Fig.~\ref{fig1}(a), we set the $z$-axis along the chiral axis and assume that the electrons are highly anisotropic to this direction, which is treated as a 1D electron system. Nevertheless, the three-dimensional features of the electrons are incorporated into the electron pseudo-angular momentum (see Fig.~\ref{fig1}(b) and Ref~\cite{SM1}).

Chirality, including static and dynamical properties, can be defined by the electric toroidal monopole $G_0$, which consists of time-reversal ($\mathcal{T}$)-even pseudoscalars~\cite{Barron2004,Kusunose2024AppPhysLett,Oiwa2022PRL}. The energy dispersion of truly chiral phonons splits due to the structural chirality during propagation along the chiral axis. The phonon frequency $\omega_{q,\lambda}$, where $q$ is the wave vector along the $z$-axis and $\lambda=\pm$ is the handedness of the anti-clockwise/clockwise-circularly polarized phonon, satisfies the relations $\omega_{q,\lambda} \neq\omega_{-q,\lambda}$ and $\omega_{q,\lambda}=\omega_{-q,-\lambda}$~\cite{Ishito2023NatPhys,Ueda2023Nature,Chen2022NanoLett,Kishine2020PRL}. The former denotes the inversion ($\mathcal{P}$) symmetry breaking, while the latter is due to the preservation of the $\mathcal{T}$ symmetry.

We here consider the longitudinal acoustic (LA) mode and transverse acoustic (TA) modes. The index of these modes is labeled $\eta=(\mathrm{L},+,-)\,[\bar{\eta}=(\mathrm{L},-,+)]$ in a unified manner. The TA modes possess phonon AM $l_{z,\pm}^{\mathrm{ph}}=\pm1$ at each $q$, while the LA mode does not. The displacement vector is written in the continuum limit as
\begin{align}
    \mathbf{u}(\bm{r})=&\sum_{q,\eta} \sqrt{\dfrac{\hbar}{2N_{\mathrm{i}}M_{\mathrm{i}}\omega_{q,\eta}}}\bm{\epsilon}_{q,\eta}(\hat{b}_{q,\eta}+\hat{b}^{\dagger}_{-q,\bar{\eta}})e^{iq z} \notag \\
    \equiv&\dfrac{1}{\sqrt{N_{\mathrm{i}}}} \sum_{q,\eta} \bm{\epsilon}_{q,\eta}\hat{\zeta}_{q,\eta}e^{iqz},
\end{align}
where $\hat{b}_{q,\eta}(\hat{b}_{q,\eta}^{\dagger})$ is the phonon annihilation (creation) operator, $\hat{\zeta}_{q,\eta}$ is the phonon field operator, and $M_{\mathrm{i}}$ is the mass of ions. Polarization vectors are defined as $\bm{\epsilon}_{q,\mathrm{L}}=(0,0,1)^{T}$ and $\bm{\epsilon}_{q,\lambda}=\frac{1}{\sqrt{2}}(1,\lambda i,0)^T$. 

In the conventional Peierls transition, only the LA phonon is considered based on the basic EPC. However, in previous research~\cite{Luo2023PRX}, the EPC for the LA mode and the TA modes was derived using the Bir-Pikus formalism~\cite{Bir&Pikus1974}, where rotational symmetry imposes selection rules on the EPC for both LA and TA phonons. Here, we extend this concept to our system in the presence of the screw symmetry. The minimal EPC Hamiltonian for each mode is given by~\cite{SM1}
\begin{align}
    \hat{H}^{(\mathrm{L})}_{\mathrm{ep}}=&\dfrac{1}{\sqrt{N_{\mathrm{i}}}}\sum_{q}iqg_{\mathrm{L}}\hat{S}^{(0)}_{-q}\hat{\zeta}_{q,\mathrm{L}},
\end{align}
\begin{align}
    \hat{H}^{(\mathrm{T})}_{\mathrm{ep}}=&\dfrac{1}{\sqrt{N_{\mathrm{i}}}}\sum_{q}iqg_{\mathrm{T}}[\hat{S}^{(+)}_{-q}\hat{\zeta}_{q,+}+\hat{S}^{(-)}_{-q}\hat{\zeta}_{q,-}],
\end{align}
where $g_{\mathrm{L}}$ and $g_{\mathrm{T}}$ denote the coupling constants and $\hat{S}_{-q}^{(X)}=\sum_{k,s,s'}\hat{c}^{\dagger}_{k+q,s}(\hat{\sigma}_{X})_{ss'}\hat{c}_{k,s'}\,\,[X=0,z,\pm]$ indicates the density wave operator, with electron momentum $k$, spin $s=\uparrow, \,\downarrow$, and Pauli matrices $\hat{\sigma}_X$. The global spin quantization axis is chosen along the $z$-axis. The screw symmetry gives the selection rule on the EPC $s = s'+l_{z,\lambda}^{\mathrm{ph}}$ for TA phonons $\hat{H}^{(\mathrm{T})}_{\mathrm{ep}}$ and $s = s'$ for LA phonons $\hat{H}^{(\mathrm{L})}_{\mathrm{ep}}$. 

We begin with the following total Hamiltonian, which includes the electron and phonon parts:
$\hat{H}=\hat{H}_{\mathrm{el}} + \hat{H}_{\mathrm{ph}} + \hat{H}_{\mathrm{ep}}$. The EPC part is given by $\hat{H}_{\mathrm{ep}} = \hat{H}^{(\mathrm{L})}_{\mathrm{ep}} + \hat{H}^{(\mathrm{T})}_{\mathrm{ep}}$. To capture the essential role of chiral phonons, we assume spin-degenerate electronic bands:
$\hat{H}_{\mathrm{el}}=\sum_{k,s}\xi_{k,s}\hat{c}^{\dagger}_{k,s}\hat{c}_{k,s}$, where $\xi_{k,s}=\hbar^2k^2/2m_e-\varepsilon_{\mathrm{F}}$ is the electron dispersion measured from the Fermi level within the parabolic approximation.

\begin{figure*}[t!]
    \centering
    \includegraphics[width=0.9\linewidth]{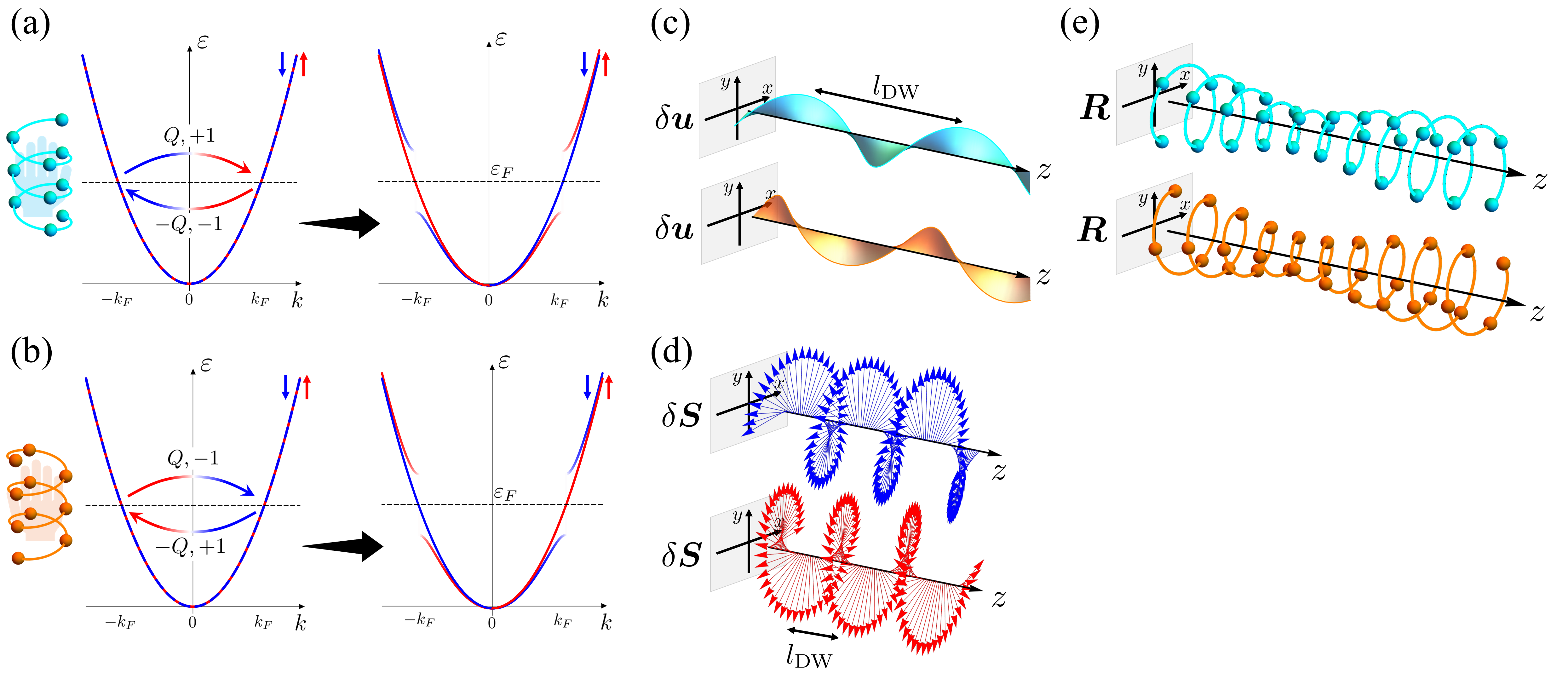}
    \caption{Incommensurate CISSPT. (a),(b) Spin-selective gap opening in the electronic band structure for the LH and RH crystals. The bands are spin degenerate above the transition temperature (left), while a spin-selective gap opens below it (right). The red-blue gradient arrows indicate the nesting vectors associated with phonon AM. (c) Chiral lattice distortion in the LH (top) and RH (bottom) crystals. (d) Helical SDW in the LH (top) and RH (bottom) crystals, with wavelength $l_{\mathrm{DW}}=2\pi/Q$. (e) The entire crystal geometry $\bm{R}_0+\delta\bm{u}(\bm{R}_0)$ below the transition temperature.}
    \label{fig2}
\end{figure*}

\textit{Kohn anomaly of chiral phonons}.---
We first discuss the renormalization of chiral phonons by the EPC, a phenomenon commonly referred to as the Kohn anomaly. This effect leads to the softening of the relevant phonon branch in 1D systems. The renormalized phonon frequencies for each branch are given by~\cite{SM2}
\begin{align}
    \hbar\tilde{\omega}_{q,\mathrm{L}}=&\hbar\omega_{q,\mathrm{L}}\sqrt{1-\dfrac{2g_{\mathrm{L}}^2q^2}{M_{\mathrm{i}} \omega_{q,\mathrm{L}}^2}\mathcal{L}(q,0)} ,
    \\
    \hbar\tilde{\omega}_{q,\lambda}=&\hbar\omega_{q,\lambda}\sqrt{1-\dfrac{4g_{\mathrm{T}}^2q^2}{M_{\mathrm{i}} \omega_{q,\lambda}^2}\mathcal{L}(q,0)}\,\,\,\,[\lambda=\pm],
\end{align}
where $\mathcal{L}(q,\omega)$ is the Lindhard function. At low temperatures, the logarithmic singularity of $\mathcal{L}(q,0)$ drives $\tilde{\omega}_q$ to zero at $q = \pm Q$, where $Q=2k_{\mathrm{F}}$ is twice the Fermi momentum. Thus, the transition temperature of the individual mode is calculated as 
\begin{align}
    k_{\mathrm{B}}T^{(\mathrm{LA})}=&\dfrac{2e^{\gamma}}{\pi}\varepsilon_{\mathrm{F}}\exp \left(-\dfrac{ M_{\mathrm{i}}\omega_{Q,\mathrm{L}}^2}{4k_{\mathrm{F}}^2g_{\mathrm{L}}^2D(\varepsilon_{\mathrm{F}})}\right), 
    \\
    k_{\mathrm{B}}T_{\lambda}^{(\mathrm{TA})}=&\dfrac{2e^{\gamma}}{\pi}\varepsilon_{\mathrm{F}}\exp \left(-\dfrac{M_{\mathrm{i}}\omega_{Q,\lambda}^2}{8k_{\mathrm{F}}^2g_{\mathrm{T}}^2D(\varepsilon_{\mathrm{F}})}\right)\,\,\,\,[\lambda=\pm],
    \label{eq.transition_temperature}
\end{align}
with the Euler constant $\gamma$ and the density of states at the Fermi level $D(\varepsilon_{\mathrm{F}})$. By comparing these transition temperatures, the realized phase can be inferred. When $T^{(\mathrm{LA})} > \max  \{ T^{(\mathrm{TA})}_{+},T^{(\mathrm{TA})}_{-}\}$, the predominant freezing of the LA mode leads to the conventional Peierls transition. In the other cases, the TA chiral phonons are preferentially frozen, and we hereafter focus on the latter case.

Furthermore, the inequivalence between $T_{+}^{(\mathrm{TA})}$ and $T_{-}^{(\mathrm{TA})}$ results in a handedness-dependent phonon softening. According to Eq.~\eqref{eq.transition_temperature}, the chiral phonon mode with the lowest frequency softens. Therefore, the soft mode is a time-reversal pair of the modes, $(Q,\lambda)$ and $(-Q,-\lambda)$. Figs.~\ref{fig1}(c) and (d) show schematic illustrations of phonon dispersion at $T^{(\mathrm{TA})}_{+}$ in LH crystals and at $T^{(\mathrm{TA})}_{-}$ in RH crystals, respectively. These figures highlight the one-to-one correspondence between the handedness of the soft phonon $\lambda$ and that of the parent chiral crystal, defined by $\chi$.

\textit{Spin-selective Peierls transition}.---
Owing to this handedness-dependent signature, the properties of phonon AM $l_{z,\pm}^{\mathrm{ph}}$ as well as the linear momentum $Q$ are embedded in the soft mode. This information is subsequently transferred to the electron system through the EPC, giving rise to CISSPT. We mainly focus on the case of incommensurate $Q$, where the lattice constant $c$ and the wavelength $l_{\mathrm{DW}}=2\pi/Q$ form an irrational fraction. Throughout the main text, we consider a case with soft modes $(Q,+)$ and $(-Q,-)$ without loss of generality. 

Within the mean-field approximation, the Hamiltonian related to the electronic states becomes
\begin{align}
    \hat{H}_{\mathrm{el}}^{\mathrm{MF}}=&\sum_{k>0}\hat{\bm{C}}_k^{\dagger}\bm{H}_k\hat{\bm{C}}_k,
    \\
    \hat{\bm{C}}_k^{\dagger}\equiv&
    \begin{bmatrix}
        \hat{c}^{\dagger}_{k,\uparrow}&\hat{c}^{\dagger}_{k-Q,\downarrow}&\hat{c}^{\dagger}_{k,\downarrow}&\hat{c}^{\dagger}_{k-Q,\uparrow}
    \end{bmatrix},
    \\
    \bm{H}_k\equiv&
    \begin{bmatrix}
        \xi_{k,\uparrow} &  \Delta & 0 & 0 \\
          \Delta^* & \xi_{k-Q,\downarrow} & 0 & 0 \\
         0 & 0 & \xi_{k,\downarrow} & 0 \\
         0 & 0 & 0 & \xi_{k-Q,\uparrow}
    \end{bmatrix},
\end{align}
where $\Delta=2iQg_{\mathrm{T}}\langle\hat{\zeta}_{Q,+}\rangle/\sqrt{N_{\mathrm{i}}}$ is defined as the order parameter. Unlike the conventional Peierls transition, electronic states with opposite spins connected by the wave number $Q$ are coupled according to the selection rule $s=s'+l_{z,\lambda}^{\mathrm{ph}}$ for the EPC. Moreover, this selection rule forbids other spin-flip processes, thereby enforcing the spin-selective gap, shown in Figs.~\ref{fig2}(a) and (b). We adopt the term “spin-selective Peierls transition” following previous research~\cite{Braunecker2010PRB}, but note that, in contrast to their analogical usage, the CISSPT formulated here denotes a qualitatively distinct phenomenon in the rigorous physical sense. Along with this spin-selective gap, only the off-diagonal components in spin space $\langle\hat{S}^{(+)}_{-Q}\rangle$ ($=\langle\hat{S}^{(-)}_{Q}\rangle^*$) take finite values, whereas only diagonal components $\langle\hat{S}^{(0)}_{-Q}\rangle$ or $\langle\hat{S}^{(z)}_{-Q}\rangle$ become non-zero in conventional CDW or spin density wave (SDW) orders. As a result, the itinerant electrons themselves compose a helical SDW, shown in Fig.~\ref{fig2}(d). In CISSPT, the chiral phonon acts as a helical Overhauser field~\cite{Braunecker2010PRB,Overhauser1960PRL}, and the helical SDW is represented in real space by
\begin{align}
    \delta \mathbf{S}(\bm{r})= \dfrac{|\Delta|}{2\pi\hbar v_{\mathrm{F}}\Lambda}
    \begin{bmatrix}
       \cos(Qz+\phi) \\
       -\sin(Qz+\phi) \\
       0
    \end{bmatrix},
\end{align}
where $|\Delta|$ and $\phi$ denote the amplitude and phase of the order parameter, respectively, and $\Lambda$ is the dimensionless EPC constant~\cite{SM2}.

At the same time, the crystal undergoes a structural phase transition, and a chiral lattice distortion arises, as shown in Fig.~\ref{fig2}(c). Due to the non-zero thermal average $\langle\hat{\zeta}_{Q,+}\rangle$, the periodicity of the daughter phase is modulated in quadrature:
\begin{align}
    \delta\mathbf{u}(\bm{r})
    =\dfrac{|\Delta|}{\sqrt{2}Qg_{\mathrm{T}}}
    \begin{bmatrix}
        \cos(Qz+\phi-\pi/2) \\
        -\sin(Qz+\phi-\pi/2) \\
        0
    \end{bmatrix}.\label{eq.lattice_distort}
\end{align}
Therefore, the lattice positions shift from the original $\mathbf{R}_0$ to $\mathbf{R}=\mathbf{R}_0+\delta\mathbf{u}(\mathbf{R}_0)$ and, as a consequence, the screw symmetry is broken. The handedness of $\delta \bm{S}(\bm{r})$ and $\delta\bm{u}(\bm{r})$ is inverted in the opposite enantiomorph.

The above discussion can be straightforwardly extended to systems with spin-split electronic bands. For instance, when the spin-orbit coupling (SOC) is included in the electron Hamiltonian $\xi_{k,s}=\hbar^2k^2/2m_e+\alpha_{\mathrm{SO}}k(\hat{\sigma}_z)_{ss}-\varepsilon_{\mathrm{F}}$, the same results are obtained by treating the system as spin subbands~\cite{SM2}. This framework remains applicable to various forms of SOC, provided that spin remains a well-defined quantum number at the Fermi level~\cite{Meng2013PRB}. Furthermore, correlation effects are known to stabilize this spin-selective gap~\cite{Braunecker2010PRB}.

Note that the spin-selective gap is forbidden when $Q$ is commensurate. In this case, the nesting vectors $(Q,+)$ and $(Q,-)$ coexist in the same electronic state, and the two spin sectors couple identically, resulting in apparently conventional Peierls gaps. See more details in Ref~\cite{SM2}.

\textit{Sliding modes}.---
Finally, we discuss the collective dynamics that reflect the distinctive nature of the CISSPT. The primary excitations are phase (phason) and amplitude (amplitudon) fluctuations of the density wave; the former being the Goldstone mode underlying the sliding motion. The energy dispersion of each mode takes the same form as in conventional CDWs~\cite{SM3}:
\begin{align}
    \Omega_{\mathrm{P}}^2(q)&=\dfrac{m_e}{m_e^*}v_{\mathrm{F}}^2q^2,
    \label{eq:phason}
    \\
    \Omega_{\mathrm{A}}^2(q)&=\Lambda\omega_{Q,+}^2+\dfrac{1}{3}\dfrac{m_e}{m_e^*}v_{\mathrm{F}}^2q^2,
    \label{eq:amplitudon}
    \\
    m_e^* &\equiv m_e\left(1+\dfrac{4|\Delta|^2}{\Lambda(\hbar\omega_{Q,+})^2}\right),
\end{align}
for the phason (P) and amplitudon (A).
These behaviors arise from the coupling between electrons and chiral phonons, indicating that, unlike conventional SDWs~\cite{Gruner1994RevModPhys}, the sliding mode of the density wave entails the circular motion of ordered ions. It is also emphasized that the nature of the sliding mode is different from that of conventional CDWs~\cite{Gruner1988RevModPhys}, since the CISSPT forms a helical SDW.

Under a static driving force such as a dc electric field $E_z$, the phase of the density wave $\phi$ evolves over time and generates a drift velocity $v_d=Q^{-1}\partial\phi/\partial t\propto E_z$, which drives the sliding conduction. Without pinning, the sliding helical SDW pumps charge and spin supercurrent along the chiral axis~\cite{Kurebayashi2022PRB}. Even in the presence of pinning by impurities or defects in realistic materials, sliding can occur above a threshold field. Alongside, the structural change of atomic positions $\mathbf{R}$ transports mechanical AM on a macroscopic scale~\cite{SM3}:
\begin{align}
    \langle\bm{J}_{\mathrm{AM}}\rangle=-\dfrac{|\Delta|^2M_{\mathrm{i}}N_{\mathrm{i}}}{2Q^2g_{\mathrm{T}}^2}\langle\dot{\phi}\rangle\,\mathbf{e}_z,
\end{align}
where $\langle\cdots\rangle$ means the temporal average.
This mechanical AM superficially resembles that of chiral phonons since both involve circular ionic motions. However, unlike chiral phonons, the AM can be transported directly by an electric field. This is because the electric field directly induces a phase evolution of the helical SDW, thereby dragging the associated ordered state of lattice displacement. This mechanism renders the transport electrically switchable, providing direct control via an applied field and opening a route to harness angular momentum as a robust carrier for information and actuation beyond excitation-driven phononics.

\begin{figure}[t!]
    \centering
    \includegraphics[width=\linewidth]{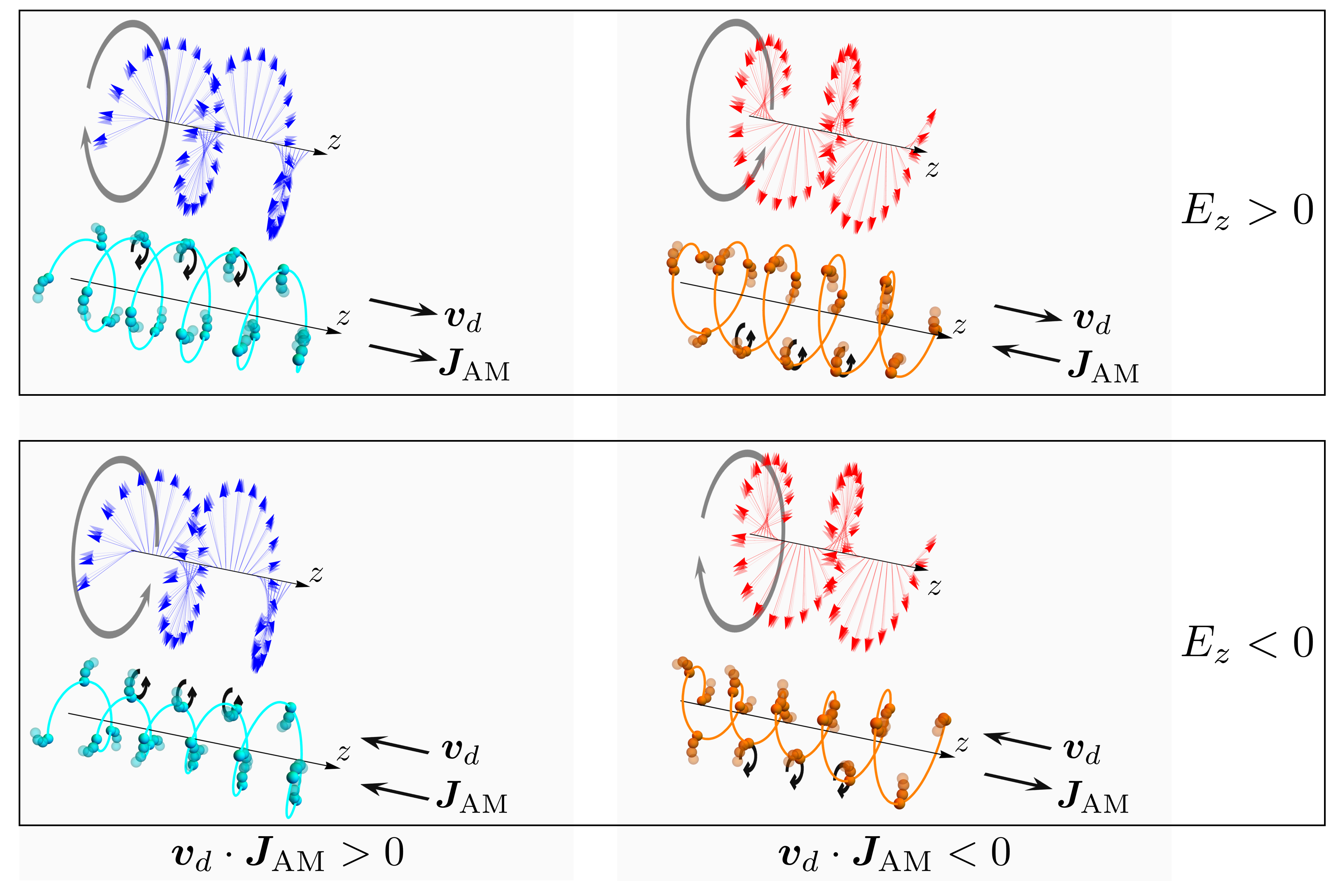}
    \caption{Sliding motion under an electric field. Upper (lower) panels show the sliding helical SDW and the accompanying lattice distortion for an electric field parallel (antiparallel) to the $z$ axis. Left and right panels correspond to LH and RH crystals, respectively. Here, $\bm{v}_{d}=v_d\,\mathbf{e}_z$ is the sliding velocity and $\bm{J}_{\mathrm{AM}}$ is the induced mechanical angular momentum. Gray arrows indicate the rotational direction of the helical SDW, while black arrows indicate the local circular motion of the lattice displacement.
}
    \label{fig3}
\end{figure}

Fig.~\ref{fig3} illustrates  
the direction of sliding velocity $\langle \bm{v}_d\rangle$ and AM $\langle \bm{J}_{\mathrm{AM}}\rangle$ depending on the handedness of crystals and the sign of the external electric field. These patterns can be classified from a multipole perspective by defining the pseudo-scalar $\bm{v}_d\cdot\bm{J}_{\mathrm{AM}}$, enabling a unified understanding of the response handedness in a similar manner to the handedness of static orders~\cite{Kusunose2024AppPhysLett}.

Beyond dc driving, the application of an ac field can trigger unique interference phenomena. When the external field frequency $\omega_{\mathrm{ac}}$ is near the characteristic pinning frequency, the phase shift is locked to $\langle\dot{\phi}\rangle=n\omega_{\mathrm{ac}}$, where $n$ is an integer. Consequently, the AM $\langle\bm{J}_{\mathrm{AM}}\rangle$ that is proportional to $\langle\dot{\phi}\rangle$  
is also locked to integer multiples in sliding mode, as well as the electric and spin current. This phenomenon is expected to be observed as plateaux, so-called Shapiro steps on the angular momentum-voltage characteristics~\cite{Kriza1991PRL,Nikiton2021AppPhysLett}.

\textit{Discussions}.---
We have demonstrated a spontaneous structural phase transition unique to chiral crystals, driven by the coupling between electrons and chiral phonons. The CISSPT belongs to the family of transverse Peierls transitions~\cite{Luo2023PRX,Yang2025NatComm,Zhang2024npj}, but unlike Ref.~\cite{Luo2023PRX}, where an imbalance of electronic spin states selects the soft phonon mode, here the intrinsic splitting of chiral phonons plays the essential role and enables the transfer of phonon AM to the electronic system. The material conditions assumed in this work may be realized in chiral chain conductors and chiral metals (see End Matter). Our formulation also predicts a complementary instability in which spin is conserved while orbital angular momentum is not, giving rise to a chiral CDW order~\cite{SM1}. The detailed properties of this state will be discussed elsewhere.

The CISSPT is distinct from the CISS effect, which persists over a wide temperature range. Nevertheless, the spin-selective Peierls gap may provide a platform for optimized spin filtering and enhanced spin response~\cite{Braunecker2009PRL,Braunecker2010PRB,Li2015PRL}. Moreover, the helical SDW stabilized by chiral lattice dynamics suggests a route to itinerant helimagnetism in structurally chiral materials such as MnSi~\cite{Stishov2011Uspekhi}, and may enable chirality-induced electromagnetic coupling relevant to multiferroic functionalities.

More broadly, mechanisms such as the CISS effect have been discussed as possible origins of biomolecular homochirality~\cite{Bloom2024ChemRev,Ozturk2022PNAS,Naaman2022AnnualRevBiophys,Michaeli2016ChemSocRev,Budin2010AnnualRevBiophys}, highlighting the broader significance of the interplay between electrons and structural chirality. In our results, the sequence of chiral orders emerges between electronic and lattice subsystems (End Matter), providing a microscopic route by which structural chirality governs electronic order and microscopic interactions generate macroscopic asymmetry. These findings may offer a useful framework for understanding chirality-driven phenomena in condensed matter and related systems.

\textit{Acknowledgments}.---
The authors are grateful to R. Sano and K. Hara for fruitful discussions. This work was supported by JSPS KAKENHI (Grant Numbers 25KJ1482, JP22H01181, JP22H04933, JP23K22452, JP23K17353, JP24K21530, JP24H00007, JP25H01249).

\bibliography{CISSPT}

\begin{thebibliography}{60}%
\makeatletter
\providecommand \@ifxundefined [1]{%
 \@ifx{#1\undefined}
}%
\providecommand \@ifnum [1]{%
 \ifnum #1\expandafter \@firstoftwo
 \else \expandafter \@secondoftwo
 \fi
}%
\providecommand \@ifx [1]{%
 \ifx #1\expandafter \@firstoftwo
 \else \expandafter \@secondoftwo
 \fi
}%
\providecommand \natexlab [1]{#1}%
\providecommand \enquote  [1]{``#1''}%
\providecommand \bibnamefont  [1]{#1}%
\providecommand \bibfnamefont [1]{#1}%
\providecommand \citenamefont [1]{#1}%
\providecommand \href@noop [0]{\@secondoftwo}%
\providecommand \href [0]{\begingroup \@sanitize@url \@href}%
\providecommand \@href[1]{\@@startlink{#1}\@@href}%
\providecommand \@@href[1]{\endgroup#1\@@endlink}%
\providecommand \@sanitize@url [0]{\catcode `\\12\catcode `\$12\catcode
  `\&12\catcode `\#12\catcode `\^12\catcode `\_12\catcode `\%12\relax}%
\providecommand \@@startlink[1]{}%
\providecommand \@@endlink[0]{}%
\providecommand \url  [0]{\begingroup\@sanitize@url \@url }%
\providecommand \@url [1]{\endgroup\@href {#1}{\urlprefix }}%
\providecommand \urlprefix  [0]{URL }%
\providecommand \Eprint [0]{\href }%
\providecommand \doibase [0]{https://doi.org/}%
\providecommand \selectlanguage [0]{\@gobble}%
\providecommand \bibinfo  [0]{\@secondoftwo}%
\providecommand \bibfield  [0]{\@secondoftwo}%
\providecommand \translation [1]{[#1]}%
\providecommand \BibitemOpen [0]{}%
\providecommand \bibitemStop [0]{}%
\providecommand \bibitemNoStop [0]{.\EOS\space}%
\providecommand \EOS [0]{\spacefactor3000\relax}%
\providecommand \BibitemShut  [1]{\csname bibitem#1\endcsname}%
\let\auto@bib@innerbib\@empty
\bibitem [{\citenamefont {Naaman}\ \emph {et~al.}(2022)\citenamefont {Naaman},
  \citenamefont {Paltiel},\ and\ \citenamefont
  {Waldeck}}]{Naaman2022AnnualRevBiophys}%
  \BibitemOpen
  \bibfield  {author} {\bibinfo {author} {\bibfnamefont {R.}~\bibnamefont
  {Naaman}}, \bibinfo {author} {\bibfnamefont {Y.}~\bibnamefont {Paltiel}},\
  and\ \bibinfo {author} {\bibfnamefont {D.~H.}\ \bibnamefont {Waldeck}},\
  }\bibfield  {title} {\bibinfo {title} {Chiral induced spin selectivity and
  its implications for biological functions},\ }\href
  {https://doi.org/https://doi.org/10.1146/annurev-biophys-083021-070400}
  {\bibfield  {journal} {\bibinfo  {journal} {Annual Review of Biophysics}\
  }\textbf {\bibinfo {volume} {51}},\ \bibinfo {pages} {99} (\bibinfo {year}
  {2022})}\BibitemShut {NoStop}%
\bibitem [{\citenamefont {Bloom}\ \emph {et~al.}(2024)\citenamefont {Bloom},
  \citenamefont {Paltiel}, \citenamefont {Naaman},\ and\ \citenamefont
  {Waldeck}}]{Bloom2024ChemRev}%
  \BibitemOpen
  \bibfield  {author} {\bibinfo {author} {\bibfnamefont {B.~P.}\ \bibnamefont
  {Bloom}}, \bibinfo {author} {\bibfnamefont {Y.}~\bibnamefont {Paltiel}},
  \bibinfo {author} {\bibfnamefont {R.}~\bibnamefont {Naaman}},\ and\ \bibinfo
  {author} {\bibfnamefont {D.~H.}\ \bibnamefont {Waldeck}},\ }\bibfield
  {title} {\bibinfo {title} {Chiral induced spin selectivity},\ }\href
  {https://doi.org/10.1021/acs.chemrev.3c00661} {\bibfield  {journal} {\bibinfo
   {journal} {Chemical Reviews}\ }\textbf {\bibinfo {volume} {124}},\ \bibinfo
  {pages} {1950} (\bibinfo {year} {2024})}\BibitemShut {NoStop}%
\bibitem [{\citenamefont {Rikken}\ and\ \citenamefont
  {Raupach}(2000)}]{Rikken2000Nature}%
  \BibitemOpen
  \bibfield  {author} {\bibinfo {author} {\bibfnamefont {G.~L. J.~A.}\
  \bibnamefont {Rikken}}\ and\ \bibinfo {author} {\bibfnamefont
  {E.}~\bibnamefont {Raupach}},\ }\bibfield  {title} {\bibinfo {title}
  {Enantioselective magnetochiral photochemistry},\ }\href
  {https://doi.org/10.1038/35016043} {\bibfield  {journal} {\bibinfo  {journal}
  {Nature}\ }\textbf {\bibinfo {volume} {405}},\ \bibinfo {pages} {932}
  (\bibinfo {year} {2000})}\BibitemShut {NoStop}%
\bibitem [{\citenamefont {Nagaosa}\ and\ \citenamefont
  {Yanase}(2024)}]{Nagaosa2024AnnualRevCondMat}%
  \BibitemOpen
  \bibfield  {author} {\bibinfo {author} {\bibfnamefont {N.}~\bibnamefont
  {Nagaosa}}\ and\ \bibinfo {author} {\bibfnamefont {Y.}~\bibnamefont
  {Yanase}},\ }\bibfield  {title} {\bibinfo {title} {Nonreciprocal transport
  and optical phenomena in quantum materials},\ }\href
  {https://doi.org/https://doi.org/10.1146/annurev-conmatphys-032822-033734}
  {\bibfield  {journal} {\bibinfo  {journal} {Annual Review of Condensed Matter
  Physics}\ }\textbf {\bibinfo {volume} {15}},\ \bibinfo {pages} {63} (\bibinfo
  {year} {2024})}\BibitemShut {NoStop}%
\bibitem [{\citenamefont {Yan}(2024)}]{Yan2024AnnualRevMatRes}%
  \BibitemOpen
  \bibfield  {author} {\bibinfo {author} {\bibfnamefont {B.}~\bibnamefont
  {Yan}},\ }\bibfield  {title} {\bibinfo {title} {Structural chirality and
  electronic chirality in quantum materials},\ }\href
  {https://doi.org/https://doi.org/10.1146/annurev-matsci-080222-033548}
  {\bibfield  {journal} {\bibinfo  {journal} {Annual Review of Materials
  Research}\ }\textbf {\bibinfo {volume} {54}},\ \bibinfo {pages} {97}
  (\bibinfo {year} {2024})}\BibitemShut {NoStop}%
\bibitem [{\citenamefont {Yang}\ \emph {et~al.}(2021)\citenamefont {Yang},
  \citenamefont {Naaman}, \citenamefont {Paltiel},\ and\ \citenamefont
  {Parkin}}]{Yang2021NatRevChem}%
  \BibitemOpen
  \bibfield  {author} {\bibinfo {author} {\bibfnamefont {S.-H.}\ \bibnamefont
  {Yang}}, \bibinfo {author} {\bibfnamefont {R.}~\bibnamefont {Naaman}},
  \bibinfo {author} {\bibfnamefont {Y.}~\bibnamefont {Paltiel}},\ and\ \bibinfo
  {author} {\bibfnamefont {S.~S.~P.}\ \bibnamefont {Parkin}},\ }\bibfield
  {title} {\bibinfo {title} {Chiral spintronics},\ }\href
  {https://doi.org/10.1038/s42254-021-00302-9} {\bibfield  {journal} {\bibinfo
  {journal} {Nature Reviews Physics}\ }\textbf {\bibinfo {volume} {3}},\
  \bibinfo {pages} {328} (\bibinfo {year} {2021})}\BibitemShut {NoStop}%
\bibitem [{\citenamefont {Inui}\ \emph {et~al.}(2020)\citenamefont {Inui},
  \citenamefont {Aoki}, \citenamefont {Nishiue}, \citenamefont {Shiota},
  \citenamefont {Kousaka}, \citenamefont {Shishido}, \citenamefont {Hirobe},
  \citenamefont {Suda}, \citenamefont {Ohe}, \citenamefont {Kishine},
  \citenamefont {Yamamoto},\ and\ \citenamefont {Togawa}}]{Akito2020prl}%
  \BibitemOpen
  \bibfield  {author} {\bibinfo {author} {\bibfnamefont {A.}~\bibnamefont
  {Inui}}, \bibinfo {author} {\bibfnamefont {R.}~\bibnamefont {Aoki}}, \bibinfo
  {author} {\bibfnamefont {Y.}~\bibnamefont {Nishiue}}, \bibinfo {author}
  {\bibfnamefont {K.}~\bibnamefont {Shiota}}, \bibinfo {author} {\bibfnamefont
  {Y.}~\bibnamefont {Kousaka}}, \bibinfo {author} {\bibfnamefont
  {H.}~\bibnamefont {Shishido}}, \bibinfo {author} {\bibfnamefont
  {D.}~\bibnamefont {Hirobe}}, \bibinfo {author} {\bibfnamefont
  {M.}~\bibnamefont {Suda}}, \bibinfo {author} {\bibfnamefont {J.-i.}\
  \bibnamefont {Ohe}}, \bibinfo {author} {\bibfnamefont {J.-i.}\ \bibnamefont
  {Kishine}}, \bibinfo {author} {\bibfnamefont {H.~M.}\ \bibnamefont
  {Yamamoto}},\ and\ \bibinfo {author} {\bibfnamefont {Y.}~\bibnamefont
  {Togawa}},\ }\bibfield  {title} {\bibinfo {title} {Chirality-induced
  spin-polarized state of a chiral crystal
  ${\mathrm{crnb}}_{3}{\mathrm{s}}_{6}$},\ }\href
  {https://doi.org/10.1103/PhysRevLett.124.166602} {\bibfield  {journal}
  {\bibinfo  {journal} {Phys. Rev. Lett.}\ }\textbf {\bibinfo {volume} {124}},\
  \bibinfo {pages} {166602} (\bibinfo {year} {2020})}\BibitemShut {NoStop}%
\bibitem [{\citenamefont {Bousquet}\ \emph {et~al.}(2025)\citenamefont
  {Bousquet}, \citenamefont {Fava}, \citenamefont {Romestan}, \citenamefont
  {Gómez-Ortiz}, \citenamefont {McCabe},\ and\ \citenamefont
  {Romero}}]{Bousquet2025JPhys}%
  \BibitemOpen
  \bibfield  {author} {\bibinfo {author} {\bibfnamefont {E.}~\bibnamefont
  {Bousquet}}, \bibinfo {author} {\bibfnamefont {M.}~\bibnamefont {Fava}},
  \bibinfo {author} {\bibfnamefont {Z.}~\bibnamefont {Romestan}}, \bibinfo
  {author} {\bibfnamefont {F.}~\bibnamefont {Gómez-Ortiz}}, \bibinfo {author}
  {\bibfnamefont {E.~E.}\ \bibnamefont {McCabe}},\ and\ \bibinfo {author}
  {\bibfnamefont {A.~H.}\ \bibnamefont {Romero}},\ }\bibfield  {title}
  {\bibinfo {title} {Structural chirality and related properties in periodic
  inorganic solids: review and perspectives},\ }\href
  {https://doi.org/10.1088/1361-648X/adb674} {\bibfield  {journal} {\bibinfo
  {journal} {Journal of Physics: Condensed Matter}\ }\textbf {\bibinfo {volume}
  {37}},\ \bibinfo {pages} {163004} (\bibinfo {year} {2025})}\BibitemShut
  {NoStop}%
\bibitem [{\citenamefont {Hayashida}\ \emph {et~al.}(2022)\citenamefont
  {Hayashida}, \citenamefont {Kimura},\ and\ \citenamefont
  {Kimura}}]{Hayashida2022ACS}%
  \BibitemOpen
  \bibfield  {author} {\bibinfo {author} {\bibfnamefont {T.}~\bibnamefont
  {Hayashida}}, \bibinfo {author} {\bibfnamefont {K.}~\bibnamefont {Kimura}},\
  and\ \bibinfo {author} {\bibfnamefont {T.}~\bibnamefont {Kimura}},\
  }\bibfield  {title} {\bibinfo {title} {Switching crystallographic chirality
  in ba(tio)cu4(po4)4 by laser irradiation},\ }\href
  {https://doi.org/10.1021/acs.jpclett.2c00606} {\bibfield  {journal} {\bibinfo
   {journal} {The Journal of Physical Chemistry Letters}\ }\textbf {\bibinfo
  {volume} {13}},\ \bibinfo {pages} {3857} (\bibinfo {year}
  {2022})}\BibitemShut {NoStop}%
\bibitem [{\citenamefont {Romao}\ and\ \citenamefont
  {Juraschek}(2024)}]{Romao2024ACSnano}%
  \BibitemOpen
  \bibfield  {author} {\bibinfo {author} {\bibfnamefont {C.~P.}\ \bibnamefont
  {Romao}}\ and\ \bibinfo {author} {\bibfnamefont {D.~M.}\ \bibnamefont
  {Juraschek}},\ }\bibfield  {title} {\bibinfo {title} {Phonon-induced
  geometric chirality},\ }\href {https://doi.org/10.1021/acsnano.4c05978}
  {\bibfield  {journal} {\bibinfo  {journal} {ACS Nano}\ }\textbf {\bibinfo
  {volume} {18}},\ \bibinfo {pages} {29550} (\bibinfo {year}
  {2024})}\BibitemShut {NoStop}%
\bibitem [{\citenamefont {Zeng}\ \emph {et~al.}(2025)\citenamefont {Zeng},
  \citenamefont {Först}, \citenamefont {Fechner}, \citenamefont {Buzzi},
  \citenamefont {Amuah}, \citenamefont {Putzke}, \citenamefont {Moll},
  \citenamefont {Prabhakaran}, \citenamefont {Radaelli},\ and\ \citenamefont
  {Cavalleri}}]{Zheng2025Science}%
  \BibitemOpen
  \bibfield  {author} {\bibinfo {author} {\bibfnamefont {Z.}~\bibnamefont
  {Zeng}}, \bibinfo {author} {\bibfnamefont {M.}~\bibnamefont {Först}},
  \bibinfo {author} {\bibfnamefont {M.}~\bibnamefont {Fechner}}, \bibinfo
  {author} {\bibfnamefont {M.}~\bibnamefont {Buzzi}}, \bibinfo {author}
  {\bibfnamefont {E.~B.}\ \bibnamefont {Amuah}}, \bibinfo {author}
  {\bibfnamefont {C.}~\bibnamefont {Putzke}}, \bibinfo {author} {\bibfnamefont
  {P.~J.~W.}\ \bibnamefont {Moll}}, \bibinfo {author} {\bibfnamefont
  {D.}~\bibnamefont {Prabhakaran}}, \bibinfo {author} {\bibfnamefont {P.~G.}\
  \bibnamefont {Radaelli}},\ and\ \bibinfo {author} {\bibfnamefont
  {A.}~\bibnamefont {Cavalleri}},\ }\bibfield  {title} {\bibinfo {title}
  {Photo-induced chirality in a nonchiral crystal},\ }\href
  {https://doi.org/10.1126/science.adr4713} {\bibfield  {journal} {\bibinfo
  {journal} {Science}\ }\textbf {\bibinfo {volume} {387}},\ \bibinfo {pages}
  {431} (\bibinfo {year} {2025})}\BibitemShut {NoStop}%
\bibitem [{\citenamefont {Zhang}\ \emph {et~al.}(2024)\citenamefont {Zhang},
  \citenamefont {Luo},\ and\ \citenamefont {Zhang}}]{Zhang2024npj}%
  \BibitemOpen
  \bibfield  {author} {\bibinfo {author} {\bibfnamefont {S.}~\bibnamefont
  {Zhang}}, \bibinfo {author} {\bibfnamefont {K.}~\bibnamefont {Luo}},\ and\
  \bibinfo {author} {\bibfnamefont {T.}~\bibnamefont {Zhang}},\ }\bibfield
  {title} {\bibinfo {title} {Understanding chiral charge-density wave by frozen
  chiral phonon},\ }\href {https://doi.org/10.1038/s41524-024-01440-1}
  {\bibfield  {journal} {\bibinfo  {journal} {npj Computational Materials}\
  }\textbf {\bibinfo {volume} {10}},\ \bibinfo {pages} {264} (\bibinfo {year}
  {2024})}\BibitemShut {NoStop}%
\bibitem [{\citenamefont {Fava}\ \emph {et~al.}(2025)\citenamefont {Fava},
  \citenamefont {Romero},\ and\ \citenamefont {Bousquet}}]{Fava2025PRL}%
  \BibitemOpen
  \bibfield  {author} {\bibinfo {author} {\bibfnamefont {M.}~\bibnamefont
  {Fava}}, \bibinfo {author} {\bibfnamefont {A.~H.}\ \bibnamefont {Romero}},\
  and\ \bibinfo {author} {\bibfnamefont {E.}~\bibnamefont {Bousquet}},\
  }\bibfield  {title} {\bibinfo {title} {Handedness selection and hysteresis of
  chiral orders in crystals},\ }\href {https://doi.org/10.1103/c4v9-nd4b}
  {\bibfield  {journal} {\bibinfo  {journal} {Phys. Rev. Lett.}\ }\textbf
  {\bibinfo {volume} {135}},\ \bibinfo {pages} {146102} (\bibinfo {year}
  {2025})}\BibitemShut {NoStop}%
\bibitem [{\citenamefont {Zhang}\ and\ \citenamefont
  {Niu}(2014)}]{Zhang2014PRL}%
  \BibitemOpen
  \bibfield  {author} {\bibinfo {author} {\bibfnamefont {L.}~\bibnamefont
  {Zhang}}\ and\ \bibinfo {author} {\bibfnamefont {Q.}~\bibnamefont {Niu}},\
  }\bibfield  {title} {\bibinfo {title} {Angular momentum of phonons and the
  einstein--de haas effect},\ }\href
  {https://doi.org/10.1103/PhysRevLett.112.085503} {\bibfield  {journal}
  {\bibinfo  {journal} {Phys. Rev. Lett.}\ }\textbf {\bibinfo {volume} {112}},\
  \bibinfo {pages} {085503} (\bibinfo {year} {2014})}\BibitemShut {NoStop}%
\bibitem [{\citenamefont {Zhu}\ \emph {et~al.}(2018)\citenamefont {Zhu},
  \citenamefont {Yi}, \citenamefont {Li}, \citenamefont {Xiao}, \citenamefont
  {Zhang}, \citenamefont {Yang}, \citenamefont {Kaindl}, \citenamefont {Li},
  \citenamefont {Wang},\ and\ \citenamefont {Zhang}}]{Hanyu2018Science}%
  \BibitemOpen
  \bibfield  {author} {\bibinfo {author} {\bibfnamefont {H.}~\bibnamefont
  {Zhu}}, \bibinfo {author} {\bibfnamefont {J.}~\bibnamefont {Yi}}, \bibinfo
  {author} {\bibfnamefont {M.-Y.}\ \bibnamefont {Li}}, \bibinfo {author}
  {\bibfnamefont {J.}~\bibnamefont {Xiao}}, \bibinfo {author} {\bibfnamefont
  {L.}~\bibnamefont {Zhang}}, \bibinfo {author} {\bibfnamefont {C.-W.}\
  \bibnamefont {Yang}}, \bibinfo {author} {\bibfnamefont {R.~A.}\ \bibnamefont
  {Kaindl}}, \bibinfo {author} {\bibfnamefont {L.-J.}\ \bibnamefont {Li}},
  \bibinfo {author} {\bibfnamefont {Y.}~\bibnamefont {Wang}},\ and\ \bibinfo
  {author} {\bibfnamefont {X.}~\bibnamefont {Zhang}},\ }\bibfield  {title}
  {\bibinfo {title} {Observation of chiral phonons},\ }\href
  {https://doi.org/10.1126/science.aar2711} {\bibfield  {journal} {\bibinfo
  {journal} {Science}\ }\textbf {\bibinfo {volume} {359}},\ \bibinfo {pages}
  {579} (\bibinfo {year} {2018})}\BibitemShut {NoStop}%
\bibitem [{\citenamefont {Wang}\ \emph {et~al.}(2024)\citenamefont {Wang},
  \citenamefont {Sun}, \citenamefont {Li},\ and\ \citenamefont
  {Zhang}}]{Wang2024NanoLett}%
  \BibitemOpen
  \bibfield  {author} {\bibinfo {author} {\bibfnamefont {T.}~\bibnamefont
  {Wang}}, \bibinfo {author} {\bibfnamefont {H.}~\bibnamefont {Sun}}, \bibinfo
  {author} {\bibfnamefont {X.}~\bibnamefont {Li}},\ and\ \bibinfo {author}
  {\bibfnamefont {L.}~\bibnamefont {Zhang}},\ }\bibfield  {title} {\bibinfo
  {title} {Chiral phonons: Prediction, verification, and application},\ }\href
  {https://doi.org/10.1021/acs.nanolett.4c00606} {\bibfield  {journal}
  {\bibinfo  {journal} {Nano Letters}\ }\textbf {\bibinfo {volume} {24}},\
  \bibinfo {pages} {4311} (\bibinfo {year} {2024})}\BibitemShut {NoStop}%
\bibitem [{\citenamefont {Ishito}\ \emph {et~al.}(2023)\citenamefont {Ishito},
  \citenamefont {Mao}, \citenamefont {Kousaka}, \citenamefont {Togawa},
  \citenamefont {Iwasaki}, \citenamefont {Zhang}, \citenamefont {Murakami},
  \citenamefont {Kishine},\ and\ \citenamefont {Satoh}}]{Ishito2023NatPhys}%
  \BibitemOpen
  \bibfield  {author} {\bibinfo {author} {\bibfnamefont {K.}~\bibnamefont
  {Ishito}}, \bibinfo {author} {\bibfnamefont {H.}~\bibnamefont {Mao}},
  \bibinfo {author} {\bibfnamefont {Y.}~\bibnamefont {Kousaka}}, \bibinfo
  {author} {\bibfnamefont {Y.}~\bibnamefont {Togawa}}, \bibinfo {author}
  {\bibfnamefont {S.}~\bibnamefont {Iwasaki}}, \bibinfo {author} {\bibfnamefont
  {T.}~\bibnamefont {Zhang}}, \bibinfo {author} {\bibfnamefont
  {S.}~\bibnamefont {Murakami}}, \bibinfo {author} {\bibfnamefont {J.-i.}\
  \bibnamefont {Kishine}},\ and\ \bibinfo {author} {\bibfnamefont
  {T.}~\bibnamefont {Satoh}},\ }\bibfield  {title} {\bibinfo {title} {Truly
  chiral phonons in $\alpha$-hgs},\ }\href
  {https://doi.org/10.1038/s41567-022-01790-x} {\bibfield  {journal} {\bibinfo
  {journal} {Nature Physics}\ }\textbf {\bibinfo {volume} {19}},\ \bibinfo
  {pages} {35} (\bibinfo {year} {2023})}\BibitemShut {NoStop}%
\bibitem [{\citenamefont {Ueda}\ \emph {et~al.}(2023)\citenamefont {Ueda},
  \citenamefont {Garc{\'i}a-Fern{\'a}ndez}, \citenamefont {Agrestini},
  \citenamefont {Romao}, \citenamefont {van~den Brink}, \citenamefont
  {Spaldin}, \citenamefont {Zhou},\ and\ \citenamefont
  {Staub}}]{Ueda2023Nature}%
  \BibitemOpen
  \bibfield  {author} {\bibinfo {author} {\bibfnamefont {H.}~\bibnamefont
  {Ueda}}, \bibinfo {author} {\bibfnamefont {M.}~\bibnamefont
  {Garc{\'i}a-Fern{\'a}ndez}}, \bibinfo {author} {\bibfnamefont
  {S.}~\bibnamefont {Agrestini}}, \bibinfo {author} {\bibfnamefont {C.~P.}\
  \bibnamefont {Romao}}, \bibinfo {author} {\bibfnamefont {J.}~\bibnamefont
  {van~den Brink}}, \bibinfo {author} {\bibfnamefont {N.~A.}\ \bibnamefont
  {Spaldin}}, \bibinfo {author} {\bibfnamefont {K.-J.}\ \bibnamefont {Zhou}},\
  and\ \bibinfo {author} {\bibfnamefont {U.}~\bibnamefont {Staub}},\ }\bibfield
   {title} {\bibinfo {title} {Chiral phonons in quartz probed by x-rays},\
  }\href {https://doi.org/10.1038/s41586-023-06016-5} {\bibfield  {journal}
  {\bibinfo  {journal} {Nature}\ }\textbf {\bibinfo {volume} {618}},\ \bibinfo
  {pages} {946} (\bibinfo {year} {2023})}\BibitemShut {NoStop}%
\bibitem [{\citenamefont {Chen}\ \emph {et~al.}(2022)\citenamefont {Chen},
  \citenamefont {Wu}, \citenamefont {Zhu}, \citenamefont {Yang}, \citenamefont
  {Gong}, \citenamefont {Gao}, \citenamefont {Yang},\ and\ \citenamefont
  {Zhang}}]{Chen2022NanoLett}%
  \BibitemOpen
  \bibfield  {author} {\bibinfo {author} {\bibfnamefont {H.}~\bibnamefont
  {Chen}}, \bibinfo {author} {\bibfnamefont {W.}~\bibnamefont {Wu}}, \bibinfo
  {author} {\bibfnamefont {J.}~\bibnamefont {Zhu}}, \bibinfo {author}
  {\bibfnamefont {Z.}~\bibnamefont {Yang}}, \bibinfo {author} {\bibfnamefont
  {W.}~\bibnamefont {Gong}}, \bibinfo {author} {\bibfnamefont {W.}~\bibnamefont
  {Gao}}, \bibinfo {author} {\bibfnamefont {S.~A.}\ \bibnamefont {Yang}},\ and\
  \bibinfo {author} {\bibfnamefont {L.}~\bibnamefont {Zhang}},\ }\bibfield
  {title} {\bibinfo {title} {Chiral phonon diode effect in chiral crystals},\
  }\href {https://doi.org/10.1021/acs.nanolett.1c04705} {\bibfield  {journal}
  {\bibinfo  {journal} {Nano Letters}\ }\textbf {\bibinfo {volume} {22}},\
  \bibinfo {pages} {1688} (\bibinfo {year} {2022})}\BibitemShut {NoStop}%
\bibitem [{\citenamefont {Kishine}\ \emph {et~al.}(2020)\citenamefont
  {Kishine}, \citenamefont {Ovchinnikov},\ and\ \citenamefont
  {Tereshchenko}}]{Kishine2020PRL}%
  \BibitemOpen
  \bibfield  {author} {\bibinfo {author} {\bibfnamefont {J.}~\bibnamefont
  {Kishine}}, \bibinfo {author} {\bibfnamefont {A.~S.}\ \bibnamefont
  {Ovchinnikov}},\ and\ \bibinfo {author} {\bibfnamefont {A.~A.}\ \bibnamefont
  {Tereshchenko}},\ }\bibfield  {title} {\bibinfo {title} {Chirality-induced
  phonon dispersion in a noncentrosymmetric micropolar crystal},\ }\href
  {https://doi.org/10.1103/PhysRevLett.125.245302} {\bibfield  {journal}
  {\bibinfo  {journal} {Phys. Rev. Lett.}\ }\textbf {\bibinfo {volume} {125}},\
  \bibinfo {pages} {245302} (\bibinfo {year} {2020})}\BibitemShut {NoStop}%
\bibitem [{\citenamefont {Juraschek}\ \emph {et~al.}(2025)\citenamefont
  {Juraschek}, \citenamefont {Geilhufe}, \citenamefont {Zhu}, \citenamefont
  {Basini}, \citenamefont {Baum}, \citenamefont {Baydin}, \citenamefont
  {Chaudhary}, \citenamefont {Fechner}, \citenamefont {Flebus}, \citenamefont
  {Grissonnanche}, \citenamefont {Kirilyuk}, \citenamefont {Lemeshko},
  \citenamefont {Maehrlein}, \citenamefont {Mignolet}, \citenamefont
  {Murakami}, \citenamefont {Niu}, \citenamefont {Nowak}, \citenamefont
  {Romao}, \citenamefont {Rostami}, \citenamefont {Satoh}, \citenamefont
  {Spaldin}, \citenamefont {Ueda},\ and\ \citenamefont
  {Zhang}}]{Juraschek2025NatPhys}%
  \BibitemOpen
  \bibfield  {author} {\bibinfo {author} {\bibfnamefont {D.~M.}\ \bibnamefont
  {Juraschek}}, \bibinfo {author} {\bibfnamefont {R.~M.}\ \bibnamefont
  {Geilhufe}}, \bibinfo {author} {\bibfnamefont {H.}~\bibnamefont {Zhu}},
  \bibinfo {author} {\bibfnamefont {M.}~\bibnamefont {Basini}}, \bibinfo
  {author} {\bibfnamefont {P.}~\bibnamefont {Baum}}, \bibinfo {author}
  {\bibfnamefont {A.}~\bibnamefont {Baydin}}, \bibinfo {author} {\bibfnamefont
  {S.}~\bibnamefont {Chaudhary}}, \bibinfo {author} {\bibfnamefont
  {M.}~\bibnamefont {Fechner}}, \bibinfo {author} {\bibfnamefont
  {B.}~\bibnamefont {Flebus}}, \bibinfo {author} {\bibfnamefont
  {G.}~\bibnamefont {Grissonnanche}}, \bibinfo {author} {\bibfnamefont {A.~I.}\
  \bibnamefont {Kirilyuk}}, \bibinfo {author} {\bibfnamefont {M.}~\bibnamefont
  {Lemeshko}}, \bibinfo {author} {\bibfnamefont {S.~F.}\ \bibnamefont
  {Maehrlein}}, \bibinfo {author} {\bibfnamefont {M.}~\bibnamefont {Mignolet}},
  \bibinfo {author} {\bibfnamefont {S.}~\bibnamefont {Murakami}}, \bibinfo
  {author} {\bibfnamefont {Q.}~\bibnamefont {Niu}}, \bibinfo {author}
  {\bibfnamefont {U.}~\bibnamefont {Nowak}}, \bibinfo {author} {\bibfnamefont
  {C.~P.}\ \bibnamefont {Romao}}, \bibinfo {author} {\bibfnamefont
  {H.}~\bibnamefont {Rostami}}, \bibinfo {author} {\bibfnamefont
  {T.}~\bibnamefont {Satoh}}, \bibinfo {author} {\bibfnamefont {N.~A.}\
  \bibnamefont {Spaldin}}, \bibinfo {author} {\bibfnamefont {H.}~\bibnamefont
  {Ueda}},\ and\ \bibinfo {author} {\bibfnamefont {L.}~\bibnamefont {Zhang}},\
  }\bibfield  {title} {\bibinfo {title} {Chiral phonons},\ }\href
  {https://doi.org/10.1038/s41567-025-03001-9} {\bibfield  {journal} {\bibinfo
  {journal} {Nature Physics}\ }\textbf {\bibinfo {volume} {21}},\ \bibinfo
  {pages} {1532} (\bibinfo {year} {2025})}\BibitemShut {NoStop}%
\bibitem [{\citenamefont {Luo}\ and\ \citenamefont {Dai}(2023)}]{Luo2023PRX}%
  \BibitemOpen
  \bibfield  {author} {\bibinfo {author} {\bibfnamefont {K.}~\bibnamefont
  {Luo}}\ and\ \bibinfo {author} {\bibfnamefont {X.}~\bibnamefont {Dai}},\
  }\bibfield  {title} {\bibinfo {title} {Transverse peierls transition},\
  }\href {https://doi.org/10.1103/PhysRevX.13.011027} {\bibfield  {journal}
  {\bibinfo  {journal} {Phys. Rev. X}\ }\textbf {\bibinfo {volume} {13}},\
  \bibinfo {pages} {011027} (\bibinfo {year} {2023})}\BibitemShut {NoStop}%
\bibitem [{\citenamefont {Budin}\ and\ \citenamefont
  {Szostak}(2010)}]{Budin2010AnnualRevBiophys}%
  \BibitemOpen
  \bibfield  {author} {\bibinfo {author} {\bibfnamefont {I.}~\bibnamefont
  {Budin}}\ and\ \bibinfo {author} {\bibfnamefont {J.~W.}\ \bibnamefont
  {Szostak}},\ }\bibfield  {title} {\bibinfo {title} {Expanding roles for
  diverse physical phenomena during the origin of life},\ }\href
  {https://doi.org/https://doi.org/10.1146/annurev.biophys.050708.133753}
  {\bibfield  {journal} {\bibinfo  {journal} {Annual Review of Biophysics}\
  }\textbf {\bibinfo {volume} {39}},\ \bibinfo {pages} {245} (\bibinfo {year}
  {2010})}\BibitemShut {NoStop}%
\bibitem [{\citenamefont {Michaeli}\ \emph {et~al.}(2016)\citenamefont
  {Michaeli}, \citenamefont {Kantor-Uriel}, \citenamefont {Naaman},\ and\
  \citenamefont {Waldeck}}]{Michaeli2016ChemSocRev}%
  \BibitemOpen
  \bibfield  {author} {\bibinfo {author} {\bibfnamefont {K.}~\bibnamefont
  {Michaeli}}, \bibinfo {author} {\bibfnamefont {N.}~\bibnamefont
  {Kantor-Uriel}}, \bibinfo {author} {\bibfnamefont {R.}~\bibnamefont
  {Naaman}},\ and\ \bibinfo {author} {\bibfnamefont {D.~H.}\ \bibnamefont
  {Waldeck}},\ }\bibfield  {title} {\bibinfo {title} {The electron{'}s spin and
  molecular chirality – how are they related and how do they affect life
  processes?},\ }\href {https://doi.org/10.1039/C6CS00369A} {\bibfield
  {journal} {\bibinfo  {journal} {Chem. Soc. Rev.}\ }\textbf {\bibinfo {volume}
  {45}},\ \bibinfo {pages} {6478} (\bibinfo {year} {2016})}\BibitemShut
  {NoStop}%
\bibitem [{\citenamefont {Viedma}(2005)}]{Vieda2005PRL}%
  \BibitemOpen
  \bibfield  {author} {\bibinfo {author} {\bibfnamefont {C.}~\bibnamefont
  {Viedma}},\ }\bibfield  {title} {\bibinfo {title} {Chiral symmetry breaking
  during crystallization: Complete chiral purity induced by nonlinear
  autocatalysis and recycling},\ }\href
  {https://doi.org/10.1103/PhysRevLett.94.065504} {\bibfield  {journal}
  {\bibinfo  {journal} {Phys. Rev. Lett.}\ }\textbf {\bibinfo {volume} {94}},\
  \bibinfo {pages} {065504} (\bibinfo {year} {2005})}\BibitemShut {NoStop}%
\bibitem [{\citenamefont {Bailey}\ \emph {et~al.}(1998)\citenamefont {Bailey},
  \citenamefont {Chrysostomou}, \citenamefont {Hough}, \citenamefont
  {Gledhill}, \citenamefont {McCall}, \citenamefont {Clark}, \citenamefont
  {Ménard},\ and\ \citenamefont {Tamura}}]{Bailey1998Science}%
  \BibitemOpen
  \bibfield  {author} {\bibinfo {author} {\bibfnamefont {J.}~\bibnamefont
  {Bailey}}, \bibinfo {author} {\bibfnamefont {A.}~\bibnamefont
  {Chrysostomou}}, \bibinfo {author} {\bibfnamefont {J.~H.}\ \bibnamefont
  {Hough}}, \bibinfo {author} {\bibfnamefont {T.~M.}\ \bibnamefont {Gledhill}},
  \bibinfo {author} {\bibfnamefont {A.}~\bibnamefont {McCall}}, \bibinfo
  {author} {\bibfnamefont {S.}~\bibnamefont {Clark}}, \bibinfo {author}
  {\bibfnamefont {F.}~\bibnamefont {Ménard}},\ and\ \bibinfo {author}
  {\bibfnamefont {M.}~\bibnamefont {Tamura}},\ }\bibfield  {title} {\bibinfo
  {title} {Circular polarization in star- formation regions: Implications for
  biomolecular homochirality},\ }\href
  {https://doi.org/10.1126/science.281.5377.672} {\bibfield  {journal}
  {\bibinfo  {journal} {Science}\ }\textbf {\bibinfo {volume} {281}},\ \bibinfo
  {pages} {672} (\bibinfo {year} {1998})}\BibitemShut {NoStop}%
\bibitem [{\citenamefont {Takano}\ \emph {et~al.}(2007)\citenamefont {Takano},
  \citenamefont {ichi Takahashi}, \citenamefont {Kaneko}, \citenamefont
  {Marumo},\ and\ \citenamefont {Kobayashi}}]{Takano2007EPSL}%
  \BibitemOpen
  \bibfield  {author} {\bibinfo {author} {\bibfnamefont {Y.}~\bibnamefont
  {Takano}}, \bibinfo {author} {\bibfnamefont {J.}~\bibnamefont {ichi
  Takahashi}}, \bibinfo {author} {\bibfnamefont {T.}~\bibnamefont {Kaneko}},
  \bibinfo {author} {\bibfnamefont {K.}~\bibnamefont {Marumo}},\ and\ \bibinfo
  {author} {\bibfnamefont {K.}~\bibnamefont {Kobayashi}},\ }\bibfield  {title}
  {\bibinfo {title} {Asymmetric synthesis of amino acid precursors in
  interstellar complex organics by circularly polarized light},\ }\href
  {https://doi.org/https://doi.org/10.1016/j.epsl.2006.11.030} {\bibfield
  {journal} {\bibinfo  {journal} {Earth and Planetary Science Letters}\
  }\textbf {\bibinfo {volume} {254}},\ \bibinfo {pages} {106} (\bibinfo {year}
  {2007})}\BibitemShut {NoStop}%
\bibitem [{\citenamefont {Gr\"uner}(1988)}]{Gruner1988RevModPhys}%
  \BibitemOpen
  \bibfield  {author} {\bibinfo {author} {\bibfnamefont {G.}~\bibnamefont
  {Gr\"uner}},\ }\bibfield  {title} {\bibinfo {title} {The dynamics of
  charge-density waves},\ }\href {https://doi.org/10.1103/RevModPhys.60.1129}
  {\bibfield  {journal} {\bibinfo  {journal} {Rev. Mod. Phys.}\ }\textbf
  {\bibinfo {volume} {60}},\ \bibinfo {pages} {1129} (\bibinfo {year}
  {1988})}\BibitemShut {NoStop}%
\bibitem [{\citenamefont {Gr\"uner}(1994{\natexlab{a}})}]{Gruner1994}%
  \BibitemOpen
  \bibfield  {author} {\bibinfo {author} {\bibfnamefont {G.}~\bibnamefont
  {Gr\"uner}},\ }\href {https://doi.org/10.1201/9780429501012} {\emph {\bibinfo
  {title} {Density Waves In Solids}}},\ \bibinfo {edition} {1st}\ ed.\
  (\bibinfo  {publisher} {CRC Press},\ \bibinfo {year} {1994})\BibitemShut
  {NoStop}%
\bibitem [{\citenamefont {Monceau}(2012)}]{Monceau2012AdvPhys}%
  \BibitemOpen
  \bibfield  {author} {\bibinfo {author} {\bibfnamefont {P.}~\bibnamefont
  {Monceau}},\ }\bibfield  {title} {\bibinfo {title} {Electronic crystals: an
  experimental overview},\ }\href
  {https://doi.org/10.1080/00018732.2012.719674} {\bibfield  {journal}
  {\bibinfo  {journal} {Advances in Physics}\ }\textbf {\bibinfo {volume}
  {61}},\ \bibinfo {pages} {325} (\bibinfo {year} {2012})}\BibitemShut
  {NoStop}%
\bibitem [{\citenamefont
  {Gr\"uner}(1994{\natexlab{b}})}]{Gruner1994RevModPhys}%
  \BibitemOpen
  \bibfield  {author} {\bibinfo {author} {\bibfnamefont {G.}~\bibnamefont
  {Gr\"uner}},\ }\bibfield  {title} {\bibinfo {title} {The dynamics of
  spin-density waves},\ }\href {https://doi.org/10.1103/RevModPhys.66.1}
  {\bibfield  {journal} {\bibinfo  {journal} {Rev. Mod. Phys.}\ }\textbf
  {\bibinfo {volume} {66}},\ \bibinfo {pages} {1} (\bibinfo {year}
  {1994}{\natexlab{b}})}\BibitemShut {NoStop}%
\bibitem [{SM1()}]{SM1}%
  \BibitemOpen
  \href@noop {} {}\bibinfo {note} {See Supplementary Note 1 in Supplementary
  Information, for more details on the formulation.}\BibitemShut {Stop}%
\bibitem [{\citenamefont {Barron}(2004)}]{Barron2004}%
  \BibitemOpen
  \bibfield  {author} {\bibinfo {author} {\bibfnamefont {L.~D.}\ \bibnamefont
  {Barron}},\ }\href@noop {} {\emph {\bibinfo {title} {Molecular Light
  Scattering and Optical Activity}}},\ \bibinfo {edition} {2nd}\ ed.\ (\bibinfo
   {publisher} {Cambridge University Press},\ \bibinfo {year}
  {2004})\BibitemShut {NoStop}%
\bibitem [{\citenamefont {Kusunose}\ \emph {et~al.}(2024)\citenamefont
  {Kusunose}, \citenamefont {Kishine},\ and\ \citenamefont
  {Yamamoto}}]{Kusunose2024AppPhysLett}%
  \BibitemOpen
  \bibfield  {author} {\bibinfo {author} {\bibfnamefont {H.}~\bibnamefont
  {Kusunose}}, \bibinfo {author} {\bibfnamefont {J.-i.}\ \bibnamefont
  {Kishine}},\ and\ \bibinfo {author} {\bibfnamefont {H.~M.}\ \bibnamefont
  {Yamamoto}},\ }\bibfield  {title} {\bibinfo {title} {Emergence of chirality
  from electron spins, physical fields, and material-field composites},\ }\href
  {https://doi.org/10.1063/5.0214919} {\bibfield  {journal} {\bibinfo
  {journal} {Applied Physics Letters}\ }\textbf {\bibinfo {volume} {124}},\
  \bibinfo {pages} {260501} (\bibinfo {year} {2024})}\BibitemShut {NoStop}%
\bibitem [{\citenamefont {Oiwa}\ and\ \citenamefont
  {Kusunose}(2022)}]{Oiwa2022PRL}%
  \BibitemOpen
  \bibfield  {author} {\bibinfo {author} {\bibfnamefont {R.}~\bibnamefont
  {Oiwa}}\ and\ \bibinfo {author} {\bibfnamefont {H.}~\bibnamefont
  {Kusunose}},\ }\bibfield  {title} {\bibinfo {title} {Rotation, electric-field
  responses, and absolute enantioselection in chiral crystals},\ }\href
  {https://doi.org/10.1103/PhysRevLett.129.116401} {\bibfield  {journal}
  {\bibinfo  {journal} {Phys. Rev. Lett.}\ }\textbf {\bibinfo {volume} {129}},\
  \bibinfo {pages} {116401} (\bibinfo {year} {2022})}\BibitemShut {NoStop}%
\bibitem [{\citenamefont {Bir}\ and\ \citenamefont
  {Pikus}(1974)}]{Bir&Pikus1974}%
  \BibitemOpen
  \bibfield  {author} {\bibinfo {author} {\bibfnamefont {G.~L.}\ \bibnamefont
  {Bir}}\ and\ \bibinfo {author} {\bibfnamefont {G.~E.}\ \bibnamefont
  {Pikus}},\ }\href {https://cir.nii.ac.jp/crid/1130282270775839488} {\emph
  {\bibinfo {title} {Symmetry and strain-induced effects in semiconductors}}}\
  (\bibinfo  {publisher} {Wiley},\ \bibinfo {year} {1974})\BibitemShut
  {NoStop}%
\bibitem [{SM2()}]{SM2}%
  \BibitemOpen
  \href@noop {} {}\bibinfo {note} {See Supplementary Note 2 for the details of
  CISSPT.}\BibitemShut {Stop}%
\bibitem [{\citenamefont {Braunecker}\ \emph {et~al.}(2010)\citenamefont
  {Braunecker}, \citenamefont {Japaridze}, \citenamefont {Klinovaja},\ and\
  \citenamefont {Loss}}]{Braunecker2010PRB}%
  \BibitemOpen
  \bibfield  {author} {\bibinfo {author} {\bibfnamefont {B.}~\bibnamefont
  {Braunecker}}, \bibinfo {author} {\bibfnamefont {G.~I.}\ \bibnamefont
  {Japaridze}}, \bibinfo {author} {\bibfnamefont {J.}~\bibnamefont
  {Klinovaja}},\ and\ \bibinfo {author} {\bibfnamefont {D.}~\bibnamefont
  {Loss}},\ }\bibfield  {title} {\bibinfo {title} {Spin-selective peierls
  transition in interacting one-dimensional conductors with spin-orbit
  interaction},\ }\href {https://doi.org/10.1103/PhysRevB.82.045127} {\bibfield
   {journal} {\bibinfo  {journal} {Phys. Rev. B}\ }\textbf {\bibinfo {volume}
  {82}},\ \bibinfo {pages} {045127} (\bibinfo {year} {2010})}\BibitemShut
  {NoStop}%
\bibitem [{\citenamefont {Overhauser}(1960)}]{Overhauser1960PRL}%
  \BibitemOpen
  \bibfield  {author} {\bibinfo {author} {\bibfnamefont {A.~W.}\ \bibnamefont
  {Overhauser}},\ }\bibfield  {title} {\bibinfo {title} {Giant spin density
  waves},\ }\href {https://doi.org/10.1103/PhysRevLett.4.462} {\bibfield
  {journal} {\bibinfo  {journal} {Phys. Rev. Lett.}\ }\textbf {\bibinfo
  {volume} {4}},\ \bibinfo {pages} {462} (\bibinfo {year} {1960})}\BibitemShut
  {NoStop}%
\bibitem [{\citenamefont {Meng}\ and\ \citenamefont
  {Loss}(2013)}]{Meng2013PRB}%
  \BibitemOpen
  \bibfield  {author} {\bibinfo {author} {\bibfnamefont {T.}~\bibnamefont
  {Meng}}\ and\ \bibinfo {author} {\bibfnamefont {D.}~\bibnamefont {Loss}},\
  }\bibfield  {title} {\bibinfo {title} {Helical nuclear spin order in
  two-subband quantum wires},\ }\href
  {https://doi.org/10.1103/PhysRevB.87.235427} {\bibfield  {journal} {\bibinfo
  {journal} {Phys. Rev. B}\ }\textbf {\bibinfo {volume} {87}},\ \bibinfo
  {pages} {235427} (\bibinfo {year} {2013})}\BibitemShut {NoStop}%
\bibitem [{SM3()}]{SM3}%
  \BibitemOpen
  \href@noop {} {}\bibinfo {note} {See Supplementary Note 3 for the
  calculations of collective modes.}\BibitemShut {Stop}%
\bibitem [{\citenamefont {Kurebayashi}\ \emph {et~al.}(2022)\citenamefont
  {Kurebayashi}, \citenamefont {Liu}, \citenamefont {Masell},\ and\
  \citenamefont {Nagaosa}}]{Kurebayashi2022PRB}%
  \BibitemOpen
  \bibfield  {author} {\bibinfo {author} {\bibfnamefont {D.}~\bibnamefont
  {Kurebayashi}}, \bibinfo {author} {\bibfnamefont {Y.}~\bibnamefont {Liu}},
  \bibinfo {author} {\bibfnamefont {J.}~\bibnamefont {Masell}},\ and\ \bibinfo
  {author} {\bibfnamefont {N.}~\bibnamefont {Nagaosa}},\ }\bibfield  {title}
  {\bibinfo {title} {Theory of charge and spin pumping in atomic-scale spiral
  magnets},\ }\href {https://doi.org/10.1103/PhysRevB.106.205110} {\bibfield
  {journal} {\bibinfo  {journal} {Phys. Rev. B}\ }\textbf {\bibinfo {volume}
  {106}},\ \bibinfo {pages} {205110} (\bibinfo {year} {2022})}\BibitemShut
  {NoStop}%
\bibitem [{\citenamefont {Kriza}\ \emph {et~al.}(1991)\citenamefont {Kriza},
  \citenamefont {Quirion}, \citenamefont {Traetteberg}, \citenamefont {Kang},\
  and\ \citenamefont {J\'erome}}]{Kriza1991PRL}%
  \BibitemOpen
  \bibfield  {author} {\bibinfo {author} {\bibfnamefont {G.}~\bibnamefont
  {Kriza}}, \bibinfo {author} {\bibfnamefont {G.}~\bibnamefont {Quirion}},
  \bibinfo {author} {\bibfnamefont {O.}~\bibnamefont {Traetteberg}}, \bibinfo
  {author} {\bibfnamefont {W.}~\bibnamefont {Kang}},\ and\ \bibinfo {author}
  {\bibfnamefont {D.}~\bibnamefont {J\'erome}},\ }\bibfield  {title} {\bibinfo
  {title} {Shapiro interference in a spin-density-wave system},\ }\href
  {https://doi.org/10.1103/PhysRevLett.66.1922} {\bibfield  {journal} {\bibinfo
   {journal} {Phys. Rev. Lett.}\ }\textbf {\bibinfo {volume} {66}},\ \bibinfo
  {pages} {1922} (\bibinfo {year} {1991})}\BibitemShut {NoStop}%
\bibitem [{\citenamefont {Nikitin}\ \emph {et~al.}(2021)\citenamefont
  {Nikitin}, \citenamefont {Zybtsev}, \citenamefont {Pokrovskii},\ and\
  \citenamefont {Loginov}}]{Nikiton2021AppPhysLett}%
  \BibitemOpen
  \bibfield  {author} {\bibinfo {author} {\bibfnamefont {M.~V.}\ \bibnamefont
  {Nikitin}}, \bibinfo {author} {\bibfnamefont {S.~G.}\ \bibnamefont
  {Zybtsev}}, \bibinfo {author} {\bibfnamefont {V.~Y.}\ \bibnamefont
  {Pokrovskii}},\ and\ \bibinfo {author} {\bibfnamefont {B.~A.}\ \bibnamefont
  {Loginov}},\ }\bibfield  {title} {\bibinfo {title} {Mechanically induced
  shapiro steps: Enormous effect of vibrations on the charge-density wave
  transport},\ }\href {https://doi.org/10.1063/5.0051436} {\bibfield  {journal}
  {\bibinfo  {journal} {Applied Physics Letters}\ }\textbf {\bibinfo {volume}
  {118}},\ \bibinfo {pages} {223105} (\bibinfo {year} {2021})}\BibitemShut
  {NoStop}%
\bibitem [{\citenamefont {Yang}\ \emph {et~al.}(2025)\citenamefont {Yang},
  \citenamefont {Luo}, \citenamefont {Zhang}, \citenamefont {Guo},
  \citenamefont {Meier}, \citenamefont {Ni}, \citenamefont {Li}, \citenamefont
  {Mercado~Lozano}, \citenamefont {Fabbris}, \citenamefont {Said},
  \citenamefont {Nelson}, \citenamefont {Zhang}, \citenamefont {May},
  \citenamefont {McGuire}, \citenamefont {Juneja}, \citenamefont {Lindsay},
  \citenamefont {Lee}, \citenamefont {Zuo}, \citenamefont {Chi}, \citenamefont
  {Dai}, \citenamefont {Zhao},\ and\ \citenamefont {Miao}}]{Yang2025NatComm}%
  \BibitemOpen
  \bibfield  {author} {\bibinfo {author} {\bibfnamefont {F.~Z.}\ \bibnamefont
  {Yang}}, \bibinfo {author} {\bibfnamefont {K.~F.}\ \bibnamefont {Luo}},
  \bibinfo {author} {\bibfnamefont {W.}~\bibnamefont {Zhang}}, \bibinfo
  {author} {\bibfnamefont {X.}~\bibnamefont {Guo}}, \bibinfo {author}
  {\bibfnamefont {W.~R.}\ \bibnamefont {Meier}}, \bibinfo {author}
  {\bibfnamefont {H.}~\bibnamefont {Ni}}, \bibinfo {author} {\bibfnamefont
  {H.~X.}\ \bibnamefont {Li}}, \bibinfo {author} {\bibfnamefont
  {P.}~\bibnamefont {Mercado~Lozano}}, \bibinfo {author} {\bibfnamefont
  {G.}~\bibnamefont {Fabbris}}, \bibinfo {author} {\bibfnamefont {A.~H.}\
  \bibnamefont {Said}}, \bibinfo {author} {\bibfnamefont {C.}~\bibnamefont
  {Nelson}}, \bibinfo {author} {\bibfnamefont {T.~T.}\ \bibnamefont {Zhang}},
  \bibinfo {author} {\bibfnamefont {A.~F.}\ \bibnamefont {May}}, \bibinfo
  {author} {\bibfnamefont {M.~A.}\ \bibnamefont {McGuire}}, \bibinfo {author}
  {\bibfnamefont {R.}~\bibnamefont {Juneja}}, \bibinfo {author} {\bibfnamefont
  {L.}~\bibnamefont {Lindsay}}, \bibinfo {author} {\bibfnamefont {H.~N.}\
  \bibnamefont {Lee}}, \bibinfo {author} {\bibfnamefont {J.-M.}\ \bibnamefont
  {Zuo}}, \bibinfo {author} {\bibfnamefont {M.~F.}\ \bibnamefont {Chi}},
  \bibinfo {author} {\bibfnamefont {X.}~\bibnamefont {Dai}}, \bibinfo {author}
  {\bibfnamefont {L.}~\bibnamefont {Zhao}},\ and\ \bibinfo {author}
  {\bibfnamefont {H.}~\bibnamefont {Miao}},\ }\bibfield  {title} {\bibinfo
  {title} {Incommensurate transverse peierls transition and signature of chiral
  charge density wave in $\mathrm{EuAl}_4$},\ }\href
  {https://doi.org/10.1038/s41467-025-65374-y} {\bibfield  {journal} {\bibinfo
  {journal} {Nature Communications}\ }\textbf {\bibinfo {volume} {16}},\
  \bibinfo {pages} {10401} (\bibinfo {year} {2025})}\BibitemShut {NoStop}%
\bibitem [{\citenamefont {Braunecker}\ \emph {et~al.}(2009)\citenamefont
  {Braunecker}, \citenamefont {Simon},\ and\ \citenamefont
  {Loss}}]{Braunecker2009PRL}%
  \BibitemOpen
  \bibfield  {author} {\bibinfo {author} {\bibfnamefont {B.}~\bibnamefont
  {Braunecker}}, \bibinfo {author} {\bibfnamefont {P.}~\bibnamefont {Simon}},\
  and\ \bibinfo {author} {\bibfnamefont {D.}~\bibnamefont {Loss}},\ }\bibfield
  {title} {\bibinfo {title} {Nuclear magnetism and electronic order in
  $^{13}\mathrm{C}$ nanotubes},\ }\href
  {https://doi.org/10.1103/PhysRevLett.102.116403} {\bibfield  {journal}
  {\bibinfo  {journal} {Phys. Rev. Lett.}\ }\textbf {\bibinfo {volume} {102}},\
  \bibinfo {pages} {116403} (\bibinfo {year} {2009})}\BibitemShut {NoStop}%
\bibitem [{\citenamefont {Li}\ \emph {et~al.}(2015)\citenamefont {Li},
  \citenamefont {Wang}, \citenamefont {Fu}, \citenamefont {Du}, \citenamefont
  {Schreiber}, \citenamefont {Mu}, \citenamefont {Liu}, \citenamefont
  {Sullivan}, \citenamefont {Cs\'athy}, \citenamefont {Lin},\ and\
  \citenamefont {Du}}]{Li2015PRL}%
  \BibitemOpen
  \bibfield  {author} {\bibinfo {author} {\bibfnamefont {T.}~\bibnamefont
  {Li}}, \bibinfo {author} {\bibfnamefont {P.}~\bibnamefont {Wang}}, \bibinfo
  {author} {\bibfnamefont {H.}~\bibnamefont {Fu}}, \bibinfo {author}
  {\bibfnamefont {L.}~\bibnamefont {Du}}, \bibinfo {author} {\bibfnamefont
  {K.~A.}\ \bibnamefont {Schreiber}}, \bibinfo {author} {\bibfnamefont
  {X.}~\bibnamefont {Mu}}, \bibinfo {author} {\bibfnamefont {X.}~\bibnamefont
  {Liu}}, \bibinfo {author} {\bibfnamefont {G.}~\bibnamefont {Sullivan}},
  \bibinfo {author} {\bibfnamefont {G.~A.}\ \bibnamefont {Cs\'athy}}, \bibinfo
  {author} {\bibfnamefont {X.}~\bibnamefont {Lin}},\ and\ \bibinfo {author}
  {\bibfnamefont {R.-R.}\ \bibnamefont {Du}},\ }\bibfield  {title} {\bibinfo
  {title} {Observation of a helical luttinger liquid in
  $\mathrm{InAs}/\mathrm{GaSb}$ quantum spin hall edges},\ }\href
  {https://doi.org/10.1103/PhysRevLett.115.136804} {\bibfield  {journal}
  {\bibinfo  {journal} {Phys. Rev. Lett.}\ }\textbf {\bibinfo {volume} {115}},\
  \bibinfo {pages} {136804} (\bibinfo {year} {2015})}\BibitemShut {NoStop}%
\bibitem [{\citenamefont {Stishov}\ and\ \citenamefont
  {Petrova}(2011)}]{Stishov2011Uspekhi}%
  \BibitemOpen
  \bibfield  {author} {\bibinfo {author} {\bibfnamefont {S.~M.}\ \bibnamefont
  {Stishov}}\ and\ \bibinfo {author} {\bibfnamefont {A.~E.}\ \bibnamefont
  {Petrova}},\ }\bibfield  {title} {\bibinfo {title} {Itinerant helimagnet
  mnsi},\ }\href {https://doi.org/10.3367/UFNe.0181.201111b.1157} {\bibfield
  {journal} {\bibinfo  {journal} {Physics-Uspekhi}\ }\textbf {\bibinfo {volume}
  {54}},\ \bibinfo {pages} {1117} (\bibinfo {year} {2011})}\BibitemShut
  {NoStop}%
\bibitem [{\citenamefont {Ozturk}\ and\ \citenamefont
  {Sasselov}(2022)}]{Ozturk2022PNAS}%
  \BibitemOpen
  \bibfield  {author} {\bibinfo {author} {\bibfnamefont {S.~F.}\ \bibnamefont
  {Ozturk}}\ and\ \bibinfo {author} {\bibfnamefont {D.~D.}\ \bibnamefont
  {Sasselov}},\ }\bibfield  {title} {\bibinfo {title} {On the origins of
  life’s homochirality: Inducing enantiomeric excess with spin-polarized
  electrons},\ }\href {https://doi.org/10.1073/pnas.2204765119} {\bibfield
  {journal} {\bibinfo  {journal} {Proceedings of the National Academy of
  Sciences}\ }\textbf {\bibinfo {volume} {119}},\ \bibinfo {pages}
  {e2204765119} (\bibinfo {year} {2022})}\BibitemShut {NoStop}%
\bibitem [{SM4()}]{SM4}%
  \BibitemOpen
  \href@noop {} {}\bibinfo {note} {See Supplementary Note 4 for the details for
  End Matter.}\BibitemShut {Stop}%
\bibitem [{\citenamefont {Nakajima}\ \emph {et~al.}(2024)\citenamefont
  {Nakajima}, \citenamefont {Pop},\ and\ \citenamefont
  {Avarvari}}]{Nakajima2024RSC}%
  \BibitemOpen
  \bibfield  {author} {\bibinfo {author} {\bibfnamefont {R.}~\bibnamefont
  {Nakajima}}, \bibinfo {author} {\bibfnamefont {F.}~\bibnamefont {Pop}},\ and\
  \bibinfo {author} {\bibfnamefont {N.}~\bibnamefont {Avarvari}},\ }\bibfield
  {title} {\bibinfo {title} {Superconductors with structural chirality},\
  }\href {https://doi.org/10.1039/D4TC01719F} {\bibfield  {journal} {\bibinfo
  {journal} {J. Mater. Chem. C}\ }\textbf {\bibinfo {volume} {12}},\ \bibinfo
  {pages} {12207} (\bibinfo {year} {2024})}\BibitemShut {NoStop}%
\bibitem [{\citenamefont {Zhang}\ \emph {et~al.}(2025)\citenamefont {Zhang},
  \citenamefont {Peshcherenko}, \citenamefont {Yang}, \citenamefont {Ward},
  \citenamefont {Raghuvanshi}, \citenamefont {Lindsay}, \citenamefont {Felser},
  \citenamefont {Zhang}, \citenamefont {Yan},\ and\ \citenamefont
  {Miao}}]{Zhang2025NatPhys}%
  \BibitemOpen
  \bibfield  {author} {\bibinfo {author} {\bibfnamefont {H.}~\bibnamefont
  {Zhang}}, \bibinfo {author} {\bibfnamefont {N.}~\bibnamefont {Peshcherenko}},
  \bibinfo {author} {\bibfnamefont {F.}~\bibnamefont {Yang}}, \bibinfo {author}
  {\bibfnamefont {T.~Z.}\ \bibnamefont {Ward}}, \bibinfo {author}
  {\bibfnamefont {P.}~\bibnamefont {Raghuvanshi}}, \bibinfo {author}
  {\bibfnamefont {L.}~\bibnamefont {Lindsay}}, \bibinfo {author} {\bibfnamefont
  {C.}~\bibnamefont {Felser}}, \bibinfo {author} {\bibfnamefont
  {Y.}~\bibnamefont {Zhang}}, \bibinfo {author} {\bibfnamefont {J.-Q.}\
  \bibnamefont {Yan}},\ and\ \bibinfo {author} {\bibfnamefont {H.}~\bibnamefont
  {Miao}},\ }\bibfield  {title} {\bibinfo {title} {Measurement of phonon
  angular momentum},\ }\href {https://doi.org/10.1038/s41567-025-02952-3}
  {\bibfield  {journal} {\bibinfo  {journal} {Nature Physics}\ }\textbf
  {\bibinfo {volume} {21}},\ \bibinfo {pages} {1387} (\bibinfo {year}
  {2025})}\BibitemShut {NoStop}%
\bibitem [{\citenamefont {Hirayama}\ \emph {et~al.}(2015)\citenamefont
  {Hirayama}, \citenamefont {Okugawa}, \citenamefont {Ishibashi}, \citenamefont
  {Murakami},\ and\ \citenamefont {Miyake}}]{Hirayama2015PRL}%
  \BibitemOpen
  \bibfield  {author} {\bibinfo {author} {\bibfnamefont {M.}~\bibnamefont
  {Hirayama}}, \bibinfo {author} {\bibfnamefont {R.}~\bibnamefont {Okugawa}},
  \bibinfo {author} {\bibfnamefont {S.}~\bibnamefont {Ishibashi}}, \bibinfo
  {author} {\bibfnamefont {S.}~\bibnamefont {Murakami}},\ and\ \bibinfo
  {author} {\bibfnamefont {T.}~\bibnamefont {Miyake}},\ }\bibfield  {title}
  {\bibinfo {title} {Weyl node and spin texture in trigonal tellurium and
  selenium},\ }\href {https://doi.org/10.1103/PhysRevLett.114.206401}
  {\bibfield  {journal} {\bibinfo  {journal} {Phys. Rev. Lett.}\ }\textbf
  {\bibinfo {volume} {114}},\ \bibinfo {pages} {206401} (\bibinfo {year}
  {2015})}\BibitemShut {NoStop}%
\bibitem [{\citenamefont {Hippert}\ \emph {et~al.}(2005)\citenamefont
  {Hippert}, \citenamefont {Geissler}, \citenamefont {Hodeau}, \citenamefont
  {Lelièvre-Berna},\ and\ \citenamefont {Regnard}}]{Hippert2005}%
  \BibitemOpen
  \bibfield  {author} {\bibinfo {author} {\bibfnamefont {F.}~\bibnamefont
  {Hippert}}, \bibinfo {author} {\bibfnamefont {E.}~\bibnamefont {Geissler}},
  \bibinfo {author} {\bibfnamefont {J.}~\bibnamefont {Hodeau}}, \bibinfo
  {author} {\bibfnamefont {E.}~\bibnamefont {Lelièvre-Berna}},\ and\ \bibinfo
  {author} {\bibfnamefont {J.}~\bibnamefont {Regnard}},\ }\href
  {https://doi.org/10.1007/1-4020-3337-0} {\emph {\bibinfo {title} {Neutron and
  X-ray Spectroscopy}}}\ (\bibinfo {year} {2005})\BibitemShut {NoStop}%
\bibitem [{\citenamefont {Wang}\ \emph {et~al.}(2025)\citenamefont {Wang},
  \citenamefont {Zhou}, \citenamefont {Ren},\ and\ \citenamefont
  {Zhang}}]{Tingting2025arxiv}%
  \BibitemOpen
  \bibfield  {author} {\bibinfo {author} {\bibfnamefont {T.}~\bibnamefont
  {Wang}}, \bibinfo {author} {\bibfnamefont {J.}~\bibnamefont {Zhou}}, \bibinfo
  {author} {\bibfnamefont {Q.}~\bibnamefont {Ren}},\ and\ \bibinfo {author}
  {\bibfnamefont {L.}~\bibnamefont {Zhang}},\ }\href
  {https://arxiv.org/abs/2507.20168} {\bibinfo {title} {Inelastic neutron
  scattering for direct detection of chiral phonons}} (\bibinfo {year}
  {2025}),\ \Eprint {https://arxiv.org/abs/2507.20168} {arXiv:2507.20168
  [cond-mat.mtrl-sci]} \BibitemShut {NoStop}%
\bibitem [{\citenamefont {Sobota}\ \emph {et~al.}(2021)\citenamefont {Sobota},
  \citenamefont {He},\ and\ \citenamefont {Shen}}]{Sobota2021RevModPhys}%
  \BibitemOpen
  \bibfield  {author} {\bibinfo {author} {\bibfnamefont {J.~A.}\ \bibnamefont
  {Sobota}}, \bibinfo {author} {\bibfnamefont {Y.}~\bibnamefont {He}},\ and\
  \bibinfo {author} {\bibfnamefont {Z.-X.}\ \bibnamefont {Shen}},\ }\bibfield
  {title} {\bibinfo {title} {Angle-resolved photoemission studies of quantum
  materials},\ }\href {https://doi.org/10.1103/RevModPhys.93.025006} {\bibfield
   {journal} {\bibinfo  {journal} {Rev. Mod. Phys.}\ }\textbf {\bibinfo
  {volume} {93}},\ \bibinfo {pages} {025006} (\bibinfo {year}
  {2021})}\BibitemShut {NoStop}%
\bibitem [{\citenamefont {Togawa}\ \emph {et~al.}(2016)\citenamefont {Togawa},
  \citenamefont {Kousaka}, \citenamefont {Inoue},\ and\ \citenamefont
  {Kishine}}]{Togawa2016JPSJ}%
  \BibitemOpen
  \bibfield  {author} {\bibinfo {author} {\bibfnamefont {Y.}~\bibnamefont
  {Togawa}}, \bibinfo {author} {\bibfnamefont {Y.}~\bibnamefont {Kousaka}},
  \bibinfo {author} {\bibfnamefont {K.}~\bibnamefont {Inoue}},\ and\ \bibinfo
  {author} {\bibfnamefont {J.-i.}\ \bibnamefont {Kishine}},\ }\bibfield
  {title} {\bibinfo {title} {Symmetry, structure, and dynamics of monoaxial
  chiral magnets},\ }\href {https://doi.org/10.7566/JPSJ.85.112001} {\bibfield
  {journal} {\bibinfo  {journal} {Journal of the Physical Society of Japan}\
  }\textbf {\bibinfo {volume} {85}},\ \bibinfo {pages} {112001} (\bibinfo
  {year} {2016})}\BibitemShut {NoStop}%
\bibitem [{\citenamefont {Ishikawa}\ \emph {et~al.}(1976)\citenamefont
  {Ishikawa}, \citenamefont {Tajima}, \citenamefont {Bloch},\ and\
  \citenamefont {Roth}}]{Ishikawa1976SolidStateComm}%
  \BibitemOpen
  \bibfield  {author} {\bibinfo {author} {\bibfnamefont {Y.}~\bibnamefont
  {Ishikawa}}, \bibinfo {author} {\bibfnamefont {K.}~\bibnamefont {Tajima}},
  \bibinfo {author} {\bibfnamefont {D.}~\bibnamefont {Bloch}},\ and\ \bibinfo
  {author} {\bibfnamefont {M.}~\bibnamefont {Roth}},\ }\bibfield  {title}
  {\bibinfo {title} {Helical spin structure in manganese silicide mnsi},\
  }\href {https://doi.org/https://doi.org/10.1016/0038-1098(76)90057-0}
  {\bibfield  {journal} {\bibinfo  {journal} {Solid State Communications}\
  }\textbf {\bibinfo {volume} {19}},\ \bibinfo {pages} {525} (\bibinfo {year}
  {1976})}\BibitemShut {NoStop}%
\bibitem [{\citenamefont {Kishine}\ \emph {et~al.}(2022)\citenamefont
  {Kishine}, \citenamefont {Kusunose},\ and\ \citenamefont
  {Yamamoto}}]{Kishine2022IsraelJChme}%
  \BibitemOpen
  \bibfield  {author} {\bibinfo {author} {\bibfnamefont {J.-i.}\ \bibnamefont
  {Kishine}}, \bibinfo {author} {\bibfnamefont {H.}~\bibnamefont {Kusunose}},\
  and\ \bibinfo {author} {\bibfnamefont {H.~M.}\ \bibnamefont {Yamamoto}},\
  }\bibfield  {title} {\bibinfo {title} {On the definition of chirality and
  enantioselective fields},\ }\href
  {https://doi.org/https://doi.org/10.1002/ijch.202200049} {\bibfield
  {journal} {\bibinfo  {journal} {Israel Journal of Chemistry}\ }\textbf
  {\bibinfo {volume} {62}},\ \bibinfo {pages} {e202200049} (\bibinfo {year}
  {2022})}\BibitemShut {NoStop}%
\bibitem [{\citenamefont {Ishioka}\ \emph {et~al.}(2010)\citenamefont
  {Ishioka}, \citenamefont {Liu}, \citenamefont {Shimatake}, \citenamefont
  {Kurosawa}, \citenamefont {Ichimura}, \citenamefont {Toda}, \citenamefont
  {Oda},\ and\ \citenamefont {Tanda}}]{Ishioka2010PRL}%
  \BibitemOpen
  \bibfield  {author} {\bibinfo {author} {\bibfnamefont {J.}~\bibnamefont
  {Ishioka}}, \bibinfo {author} {\bibfnamefont {Y.~H.}\ \bibnamefont {Liu}},
  \bibinfo {author} {\bibfnamefont {K.}~\bibnamefont {Shimatake}}, \bibinfo
  {author} {\bibfnamefont {T.}~\bibnamefont {Kurosawa}}, \bibinfo {author}
  {\bibfnamefont {K.}~\bibnamefont {Ichimura}}, \bibinfo {author}
  {\bibfnamefont {Y.}~\bibnamefont {Toda}}, \bibinfo {author} {\bibfnamefont
  {M.}~\bibnamefont {Oda}},\ and\ \bibinfo {author} {\bibfnamefont
  {S.}~\bibnamefont {Tanda}},\ }\bibfield  {title} {\bibinfo {title} {Chiral
  charge-density waves},\ }\href
  {https://doi.org/10.1103/PhysRevLett.105.176401} {\bibfield  {journal}
  {\bibinfo  {journal} {Phys. Rev. Lett.}\ }\textbf {\bibinfo {volume} {105}},\
  \bibinfo {pages} {176401} (\bibinfo {year} {2010})}\BibitemShut {NoStop}%
\end{thebibliography}%

\onecolumngrid\begin{center}\textbf{\large{End Matter}}\end{center}
\twocolumngrid

\begin{figure}[b!]
    \centering
    \includegraphics[width=\linewidth]{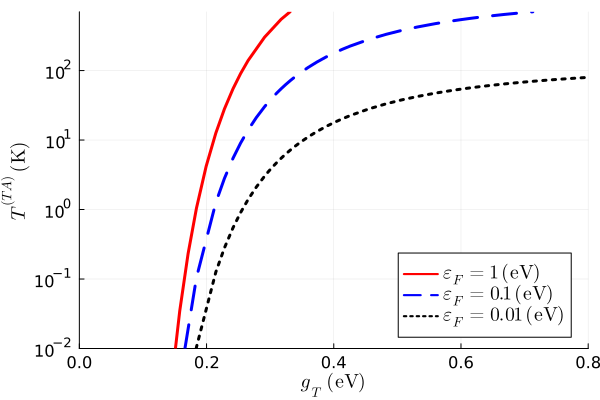}
    \caption{Transition temperature of CISSPT as a function of the coupling constant $g_{\mathrm{T}}$. $T^{(\mathrm{TA})}=\mathrm{max}\{T^{(\mathrm{TA})}_{+},T^{(\mathrm{TA})}_{-}\}$ is the transition temperature associated with the transverse Peierls instability. Curves correspond to representative Fermi energies $\varepsilon_{\mathrm{F}} = 1,\, 0.1,\, 0.01\ \mathrm{eV}$, spanning typical conduction bandwidths in realistic systems.}
    \label{fig4}
\end{figure}

\textit{Material feasibility and experimental signatures}.---
Based on the theoretical framework, the necessary conditions for CISSPT are summarized as follows: (1) the parent crystal is structurally chiral, lacking inversion and mirror symmetries while preserving screw symmetry along a principal axis; (2) the parent phase is a metal or organic conductor with itinerant carriers, providing electronic states at the Fermi level; (3) the Fermi surface exhibits nesting parallel to the chiral axis --- this requirement is not limited to quasi-1D systems; (4) the ratios $\omega_{Q,T}/g_{\mathrm{T}}$ and $\omega_{Q,L}/g_{\mathrm{L}}$ are comparable, so that the transverse instability is energetically favored.

To assess the experimental feasibility of the transverse instability, we estimate the relevant energy scales using representative rather than compound-specific parameters~\cite{SM4}. Figure~\ref{fig4} plots the transition temperature $T^{(\mathrm{TA})}$ as a function of $g_{\mathrm{T}}$ for $\varepsilon_{\mathrm{F}} = 1,\ 0.1,\ 0.01\ \mathrm{eV}$. This figure, together with Table~\ref{Table1} clarifies the range of EPC strengths necessary for CISSPT under experimentally relevant thermal conditions. The calculation indicates that CISSPT can be realized within a coupling constant regime on the order of $0.1$ eV, comparable to that of realistic materials. At this scale, the difference in transition temperatures between LH and RH phonons also becomes sufficiently large so that CISSPT naturally selects a specific phonon chirality (Table~\ref{Table2}).

These estimates demonstrate the feasibility of the Peierls instability driven by chiral phonons. The transverse Peierls transition reported in achiral EuAl$_4$~\cite{Yang2025NatComm} implies that analogous chiral systems may host CISSPT. Potential material classes span from chiral organic systems with screw symmetry, resembling $\kappa$-NCS compounds in structure and conduction~\cite{Nakajima2024RSC}, to structurally chiral elemental systems such as Te and Se, where chiral phonons have already been observed~\cite{Zhang2025NatPhys}. Although Te and Se are not metallic in their pristine forms, carrier doping or alloying may render them viable candidates~\cite{Hirayama2015PRL}.  

\begin{table}[b!]
\centering
\caption{Required transverse electron-phonon coupling $g_{\mathrm T}$ (eV) for selected transition temperatures $T^{(\mathrm{TA})}$ at representative Fermi energy scales $\varepsilon_F$.}
\setlength{\tabcolsep}{4pt}
\renewcommand{\arraystretch}{1.1}
\footnotesize
\resizebox{\columnwidth}{!}{
\begin{tabular}{c|cccc}
\hline\hline
\diagbox{$\varepsilon_{\mathrm{F}}$ (eV)}{$T^{(\mathrm{TA})}$ (K)} 
& 0.1 & 1 & 10 & 100 \\
\hline
0.01 & 0.21 & 0.26 & 0.35 & 1.0 \\
0.1  & 0.18 & 0.21 & 0.26 & 0.35 \\
1    & 0.16 & 0.18 & 0.21 & 0.26 \\
\hline
\end{tabular}
}
\label{Table1}
\end{table}

\begin{table}[b!]
\centering
\caption{Ratio $T^{(\mathrm{TA})}_{+}/T^{(\mathrm{TA})}_{-}$ for 1\% and 10\% chiral phonon splittings, $\delta\omega/\omega_{Q,\mathrm T}$ and coupling strengths $g_{\mathrm T}$.}
\setlength{\tabcolsep}{4pt}
\renewcommand{\arraystretch}{1.1}
\footnotesize
\resizebox{\columnwidth}{!}{
\begin{tabular}{c|ccccc}
\hline\hline
$g_{\mathrm{T}}$ (eV) & 0.1 & 0.2 & 0.3 & 0.4 & 0.5 \\
\hline
$\delta\omega/\omega_{Q,\mathrm{T}}=0.01$
& 0.52 & 0.85 & 0.93 & 0.96 & 0.97 \\
$\delta\omega/\omega_{Q,\mathrm{T}}=0.1$
& 0.0016 & 0.20 & 0.49 & 0.67 & 0.77 \\
\hline
\end{tabular}
}
\label{Table2}
\end{table}

\begin{figure*}[t!]
    \centering
    \includegraphics[width=0.8\linewidth]{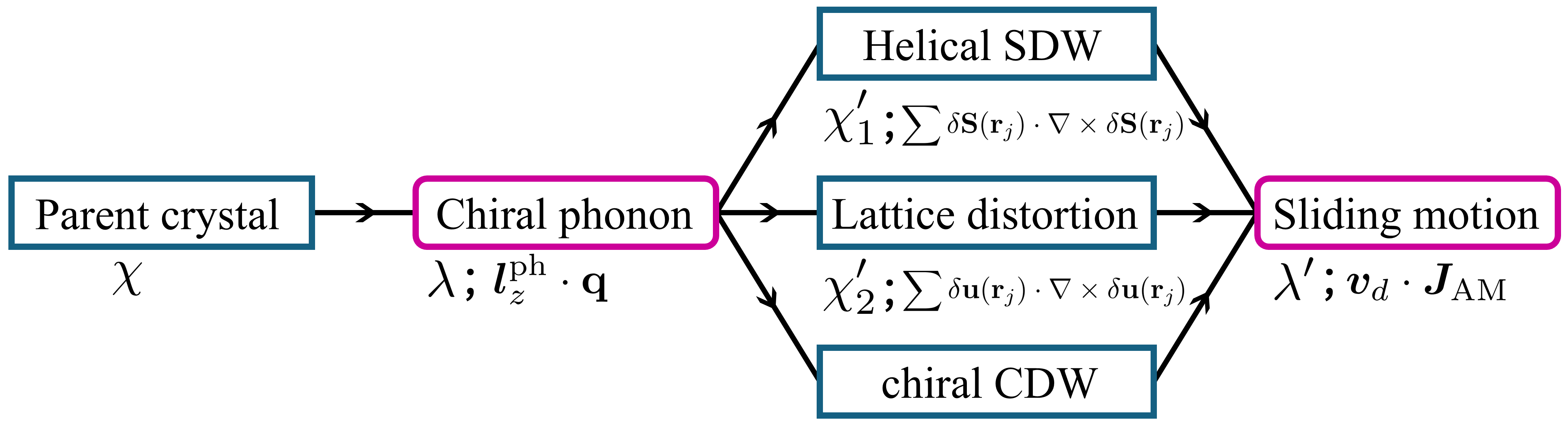}
    \caption{Flowchart representing the sequence of chiral signatures. Starting from the handedness of parental crystals, each physical quantity exhibits a cascade of one-to-one handedness correspondences. The navy rectangles represent static chirality while purple rounded rectangles represent dynamical chirality. The black arrow indicates the process by which chirality is sequentially determined.}
    \label{fig5}
\end{figure*} 

While the identification of materials hosting CISSPT remains an open challenge for future work, our work provides a unified experimental framework for identifying their signatures. Experimentally, CISSPT is identified through coupled phononic and electronic signatures. Inelastic neutron or X-ray scattering can detect the softening of acoustic phonons at finite $Q$, with distinct softening for LH and RH phonons serving as evidence of the chiral lattice instability~\cite{Hippert2005,Tingting2025arxiv}. Below the transition, a halving of conductance is expected~\cite{Braunecker2009PRL,Braunecker2010PRB,Overhauser1960PRL}, while the helical SDW and spin-selective gap can be probed by elastic neutron scattering, nuclear magnetic resonance techniques, and spin-resolved ARPES~\cite{Monceau2012AdvPhys,Gruner1994RevModPhys,Sobota2021RevModPhys}. A characteristic hallmark is the shared wave vector $Q$ of the soft phonon, density wave, and lattice modulation, together with its doping dependence, distinguishing CISSPT from conventional SOC-driven helimagnets~\cite{Togawa2016JPSJ,Ishikawa1976SolidStateComm}. 

\textit{Unified perspective of chiral signatures}.---
The phenomena predicted in this work can be understood from a unified symmetry perspective in terms of the $\mathcal{P}$-odd, $\mathcal{T}$-even pseudoscalar $G_0$, which characterizes structural chirality, collective orders, and dynamical responses~\cite{Kusunose2024AppPhysLett,Oiwa2022PRL,Kishine2022IsraelJChme}. Within this framework, the handedness of the parent crystal $\chi$ determines the handedness of the condensed chiral phonon $\lambda$, which in turn fixes the chirality of 
the emergent electronic orders $\chi'$, including the helical SDW and the accompanying chiral lattice distortion. Thus, the chiral information encoded in the lattice is transferred to the electronic subsystem through the spin-selective Peierls instability.

When the resulting density wave is driven by an electric field, the sliding motion generates a circular atomic motion carrying mechanical AM. The handedness of this dynamical motion $\lambda'$ is determined by the chirality of the underlying electronic order, which itself originates from the soft mode selected by the parent crystal. In this way, the chirality encoded in the parent lattice is converted into electronic orders through the Peierls instability and is then fed back into the lattice subsystem under the driving force.  The detailed construction of the corresponding pseudoscalars is provided in Supplementary Note 4~\cite{SM4}. Although the present analysis focuses on the helical SDW, the same symmetry correspondence can also apply to chiral CDW~\cite{SM1}: its handedness may be characterized by an appropriate pseudoscalar and is expected to be uniquely determined by the chirality of the condensed chiral phonons~\cite{Ishioka2010PRL}.

Figure~\ref{fig5} summarizes this sequential correspondence: the handedness of the parent crystal uniquely determines the chirality of the soft phonon, the phonon chirality uniquely determines the chirality of the electronic orders, and the electronic chirality in turn determines the handedness of the dynamical lattice response. The present work therefore reveals a cascade of one-to-one handedness correspondences linking structural chirality, electronic order, and driven lattice dynamics.

\newpage
\ifarXiv
  \foreach \x in {1,...,\numbersupplementpages}
  {
    \clearpage
    \includepdf[pages={\x}]{\supplementfilename}
  }
\fi

\end{document}